\documentclass[twocolumns]{aa}%

\usepackage{graphicx}
\usepackage[varg]{txfonts}
\begin{document}

   \title{Long and short term variability of seven blazars in six near-infrared/optical bands.}

   \subtitle{}

   \author{A. Sandrinelli \inst{1,2}, S. Covino \inst{2}
             \and
          A. Treves
          \inst{1,3}
  }

   \institute{Universit\`a dell'Insubria, 
              Via Valleggio 11, 22100, Como, Italy\\          
	 \and
             INAF, Osservatorio Astronomico di Brera, 
             Via Bianchi 46, 23807 Merate (LC), Italy\\
             \email{angela.sandrinelli@brera.inaf.it}
             \and
             affiliated to INAF, INFN, ICRA
             }

   \date{Received Aaaa 00, 0000; accepted Bbbb 100, 0000}

 \abstract
   {We present the light curves of six BL Lac objects, \object{PKS 0537-441}, \object{PKS 0735+17},  \object{OJ 287}, \object{PKS 2005-489}, \object{PKS 2155-304}, \object{W Comae}, and of the flat spectrum radio quasar \object{ PKS 1510-089}, as a part of a photometric monitoring  program in the near-infrared/optical bands started in 2004. All sources are Fermi blazars.}
   {Our purpose is to investigate flux and spectral variability on short and long time scales.  Systematic monitoring, independent of the activity of the source, guarantees large sample size statistics, and allows  an unbiased view of different activity states on weekly or daily time scales for the whole timeframe and on nightly timescales for some epochs.}
   { Data were obtained  with the REM telescope located at the ESO premises of La Silla (Chile).  Light curves were gathered in  the optical/near-infrared VRIJHK bands  from April 2005 to June 2012.}
   {Variability $\gtrsim$ 3 mag is observed in PKS 0537-441, PKS 1510-089 and PKS 2155-304, the largest ranges spanned in the near-infrared. The color intensity plots show rather different morphologies. The spectral energy distributions in general are well fitted by a power law, with some deviations that are more apparent  in low states. Some variability episodes during a night interval are well documented for PKS 0537-441 and PKS 2155-304. For the latter source the variability time scale implies a large relativistic beaming factor. }
   {}


\titlerunning{Variability of Seven Blazars in Six Bands}

    \maketitle{}

\section{Introduction}

Blazars are active galactic nuclei (AGNs) characterized by spectra extending from 
radio to GeV and TeV bands, high variability and polarization.  An 
important subclass is that of BL Lac objects, where contrary to other 
species of AGNs, emission lines are very weak, if not absent. Ever since 
the seminal paper of \cite{Blandford1978}, the basic model of these sources
 requires nonthermal emission from a relativistic jet, which is pointed in 
 the observer's direction. The thermal emission associated to an accretion 
 disk or to the broad emission line region is only a fraction of that from
  the jet. The jet emission is most probably dominated by synchrotron 
  radiation of relativistic electrons  and by Compton emission through 
  the scattering of electrons off synchrotron  photons or thermal ones. 
  This gives rise to the typical "two-peaked shape" of the spectral energy 
  distribution (SED), when studied over a broad energy band.
The variability, which is present at all bands, is a basic tool 
for constraining the model, since it gives information on the size 
of the emitting region and on the relativistic beaming factor, which 
transforms the quantities measured in the observer frame into those of
 the emitting region.
 
 Here we present optical-NIR photometry of seven blazars, therefore we are 
probing the synchrotron component, where the first peak of the SED is 
located. As is apparent from the literature and from our results 
\bibpunct[, ]{(}{)}{,}{a}{}{;} \citep[see e.g.][and references therein] {Impiombato2011, Bonning2012}, 
the variability pattern  of blazars in the optical is rather complex.
 On time scales of days or months, one can explore modifications in the
  jet structure or accretion disk, and possibly their interaction.  On 
  the other hand, when the time scales are hours, we are close to the 
  scales of the expected radius of the central black hole. General and
   special relativistic effects become dominant.

Long but sparse exposures have been obtained through the REM telescope, 
which being robotic is very well fitted for systematic observations of
 extensive duration. The telescope, the CCD cameras, and the photometry 
 procedures are described in section 2 and the blazar sample in section 3. 
 The variability on time scales larger than one day is presented in section 4 and  short term variability in section  5.  A discussion of the 
  results is given in section 6.

\begin{table*}
\centering
\caption{Blazar sample}
\small
\begin{tabular}{lcllccc}
\hline\hline             

   Source			& Coordinates (1)                        &  Optical           & SED		&Redshift 	&TeV Source (4)	& References \\
                           		&       RA.   DEC.                          	&  Class (2)        &  Class (3)	&		&   				& 	\\          
                           		&       [h:m:s]   [d:m:s]                   	&                         &      			&		&	   			& \\           
\hline\hline
&&&&&\\
PKS 0537-441		&    05:38:50.35 - 44:05:09.05	&      BL Lac	&  LSP		&0.896	& 				& (a)\\
PKS 0735+17		&     07:38:07.39 +17:42:18.00	&      BL Lac	&LSP		&0.424	& 				& \\
OJ 287			&   08:54.48.87 +20.06:30.64	&      BL Lac	& LSP		&0.3060	&				&(b)\\ 
PKS 1510-089		&   05:12:50.53 - 09 :05:59.83	&      FSRQ	& LSP		&0.3599	&	Y			&(c) \\
PKS 2005-489		&   20:09:25.39 - 4 8:49:53.72	&       BL Lac	& HSP		&0.071	&  	Y			&\\
PKS 2155-304		&   21:58:52.07 - 30:13:32.12 	&      BL Lac	& HSP		&  0.117	&  	Y			&(d) \\ 
W Comae 		&   12:21:31.69 + 28:13.58.50	&      BL Lac 	&ISP			& 0.1029	&    	Y	  		& (e)\\
&&&&&\\
\hline	
\end{tabular}
\tablefoot
{

(1): ICRS coordinates (J2000) and redshifts from SIMBAD\footnote{http://simbad.u-strasg.fr/}. 

(2): Class from \cite{Massaro2012}. 

(3): Classification of the spectral energy distribution: LSP means low synchrotron peaked ($\nu_S < 10^{14}$Hz, where $\nu_S$ is the synchrotron peak frequency) , ISP intermediate synchrotron peaked ($10^{14} Hz < \nu_S < 10^{15}$ Hz), and HSP high synchrotron peaked ($\nu_S > 10^{15}$ Hz) blazars;   from \cite{Abdo2010}.  

(4): TeV Sources from TeVCat\footnote{http://tevcat.unichicago.edu/}.
}
\tablebib
{
(a): \cite{Zhang2013}, \cite{Dammando2013a}, \cite{Impiombato2011}, \cite{Dammando2011a}, \cite{Dammando2010a},  \cite{Pucella2010}, \cite{Impiombato2008},  \cite{Pian2007}, \cite{Dolcini2005};

 (b): \cite{Dammando2013b};

(c): \cite{Dammando2011a},  \cite{Dammando2011b},  \cite{Dammando2011c} , \cite{Dammando2010a}, \cite{Dammando2010b},  \cite{Dammando2010c};

(d): \cite{Covino2010}, \cite{Impiombato2008},  \cite{Foschini2008}, \cite{Dolcini2007};

(e): \cite{Dammando2011a} ,  \cite{Dammando2011b}, \cite{Dammando2010a} .
}
 
\label{sample}
\end{table*}

\section{Telescope, camera and photometric procedures}

The Rapid Eye Mounting Telescope \citep[REM,][]{Zerbi2001, Covino2004} is a 
robotic telescope located at the ESO Cerro La Silla observatory (Chile). It 
was built with the main motivation to promptly observe the gamma ray burst 
detected by the \textit{Swift} mission. REM has a Ritchey-Chretien 
configuration with a 60 cm f/2.2 primary and an overall f/8 focal ratio in 
a fast-moving alt-azimuth mount that provides  two stable Nasmyth focal stations. 
The two cameras, REMIR \citep{Conconi2004} for near-IR and ROSS  
\citep{Tosti2004} for the optical, both have a field of view of 10 $\times$ 
10 arcmin and imaging capabilities with the usual NIR (z, J, H, and K) and 
Johnson-Cousins VRI filters. They allow us to obtain nearly simultaneous
data. The REM software system \citep{Covino2004} is able to manage complex 
observational strategies in a fully autonomous way.
 
 In this paper we 
consider optical and NIR data in VRIJHK bands (from 0.55 to 2.15 $\mu$m), 
collected from April 11,  2005 (53471 MJD) to June 30,  2012 (56108 MJD). 
Instrumental  magnitudes were obtained via aperture photometry, using aperture 
radii of 5 arcsec and typically with  300 s integration time in the optical 
and 150 s integration time in the NIR. Calibration was performed by means of  
comparison stars in the field reported in \textit{Two Micron All Sky Survey 
Catalog (2MASS)}\footnote{\texttt{http://www.ipac.caltech.edu/2mass/}} 
\citep{Skrutskie2006} for NIR frames. For optical standard stars, 
calibrated sequences from several authors were followed, depending on 
the blazar field.
Among the calibrated stars  we chose a \textit{reference} 
 star, present in each frame, and a \textit{check} star for each source of the sample. The check star can change depending on its positions in the frames. 
  When the check 
 star differed from its mean value by more than 2.5 $\sigma$, the image was 
 discarded. Dubious and mainly low-state images were visually  inspected in order to remove frames affected by inhomogenous background, bad tracking, etc.,  to obtain a clean photometric sample. Interesting episodes were similarly carefully inspected.

\section{The blazar sample}

 The seven blazars presented here (Table \ref{sample}) are all bright, well 
 studied objects. Among the sources 
 of the class monitored by REM\footnote{REM data are available at \texttt{http://www.rem.inaf.it}}  ($\sim$60), they were chosen with the basic 
 criterion of having  the largest extensions of the 
 total coverage. Six sources are BL Lac objects,
and one is a flat spectrum radio quasar (FSRQ). The redshift 
is known for all of them. They are all Fermi 
gamma ray sources and four of them were detected in the TeV band as shown in
 Table \ref{sample}. Recently a strong but complex correlation between optical and gamma rays 
has been demonstrated 
\citep[e.g.][]{Chatterjee2013}. Some REM photometry on the target sources
 has already been published.

\begin{figure*}
\begin{tabular}{ccc}
\includegraphics[width=0.6\columnwidth]{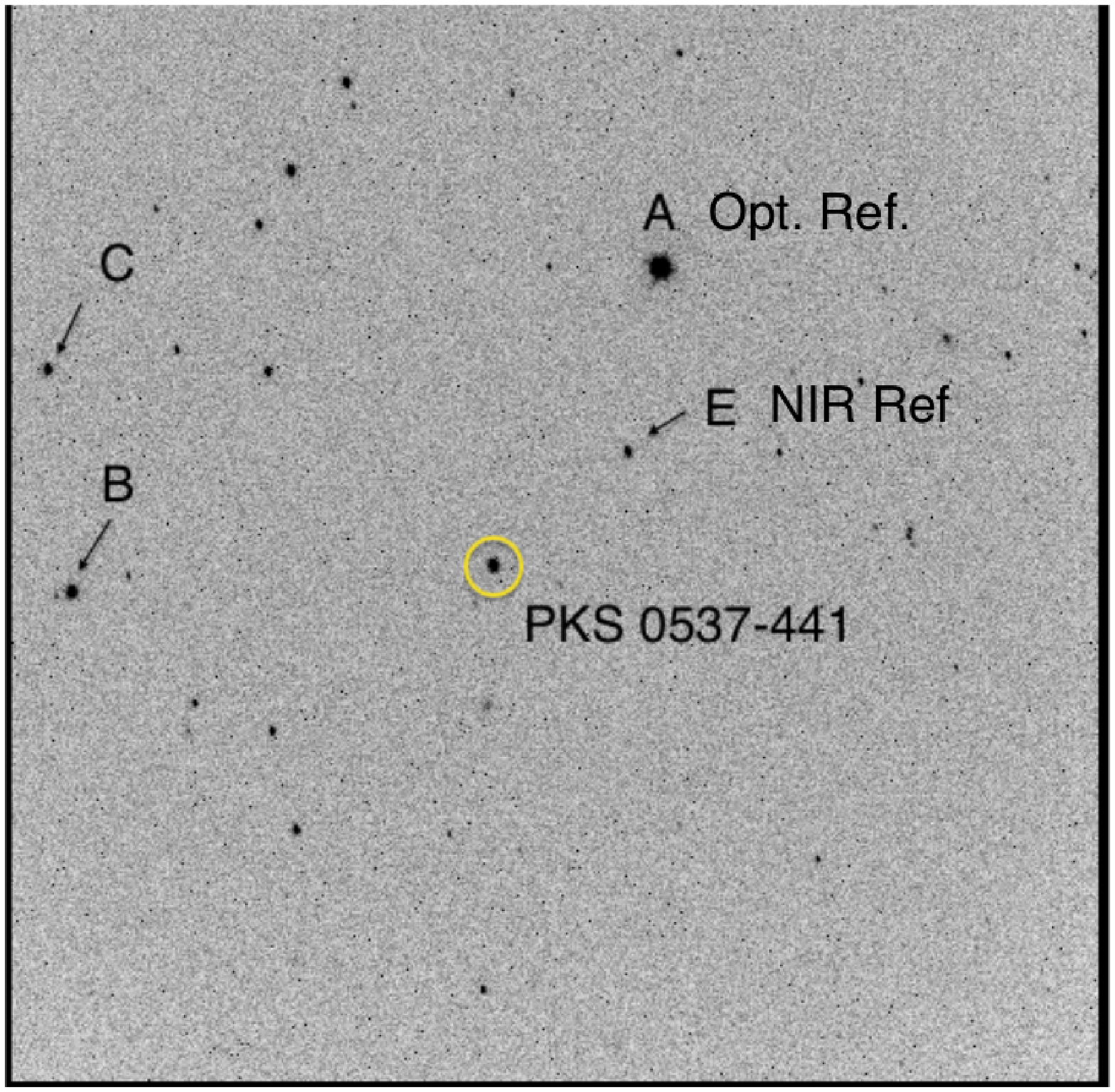}&
\includegraphics[width=0.6\columnwidth]{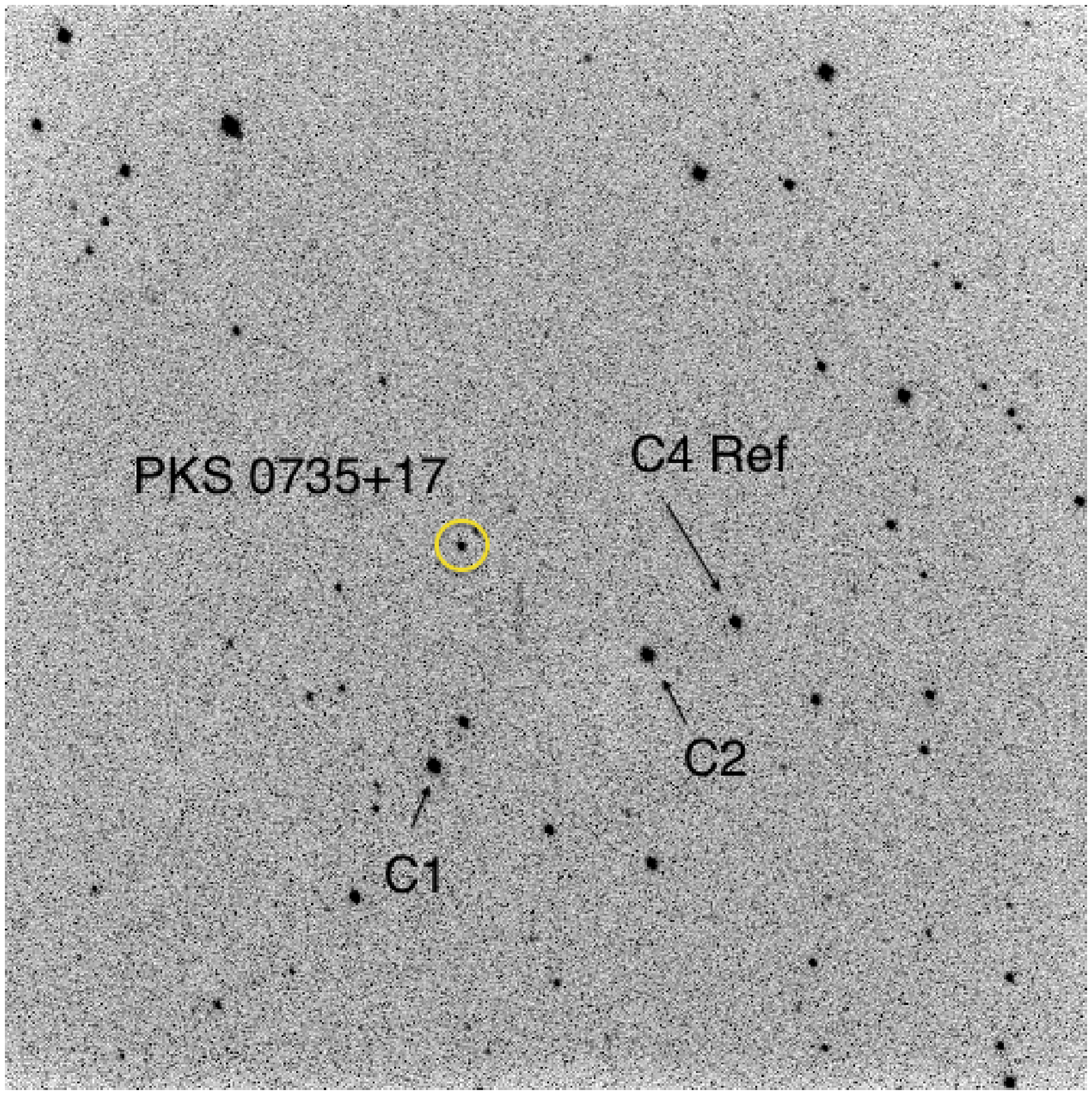}&\
\includegraphics[width=0.6\columnwidth]{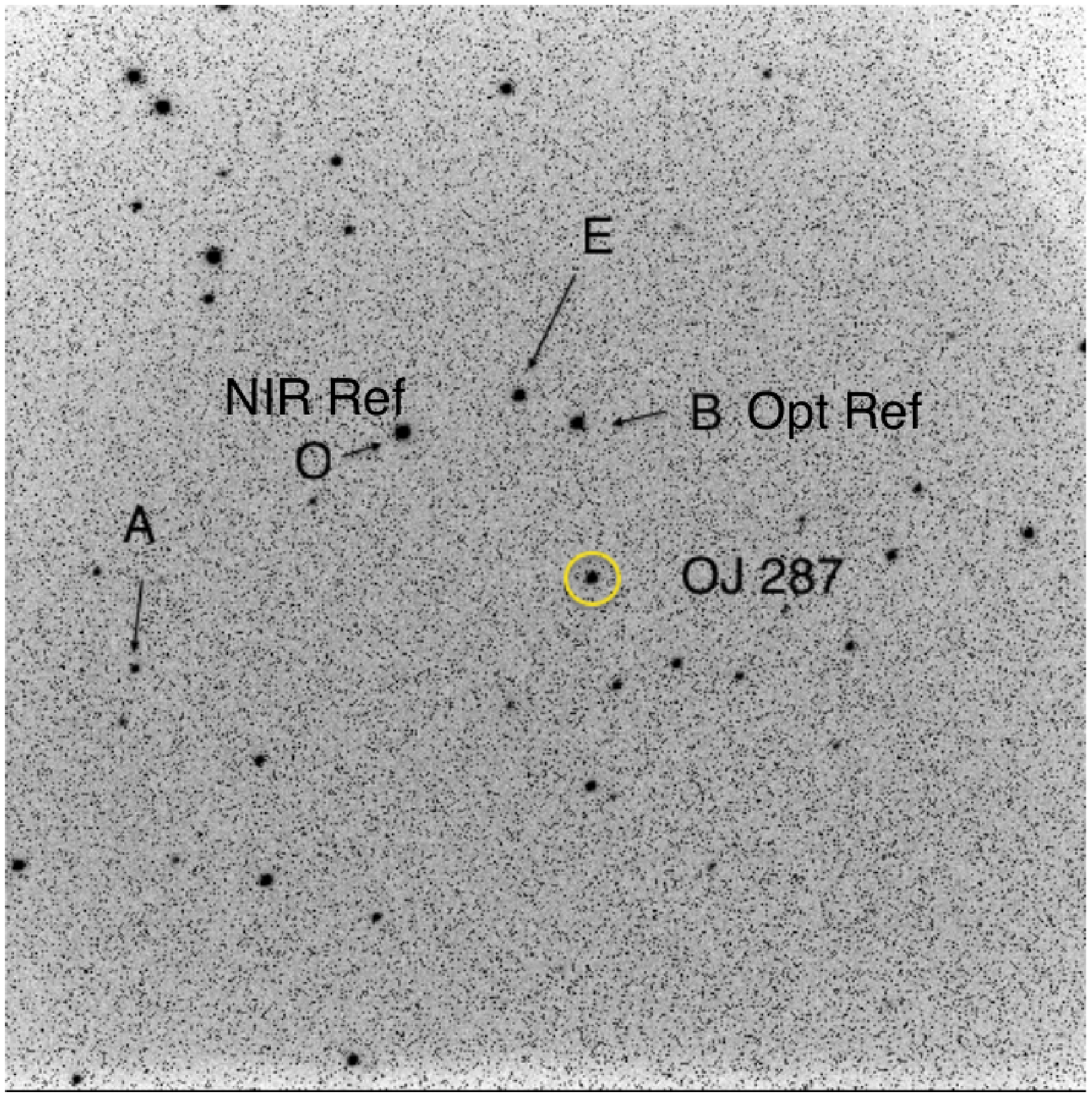}\\
\includegraphics[width=0.6\columnwidth]{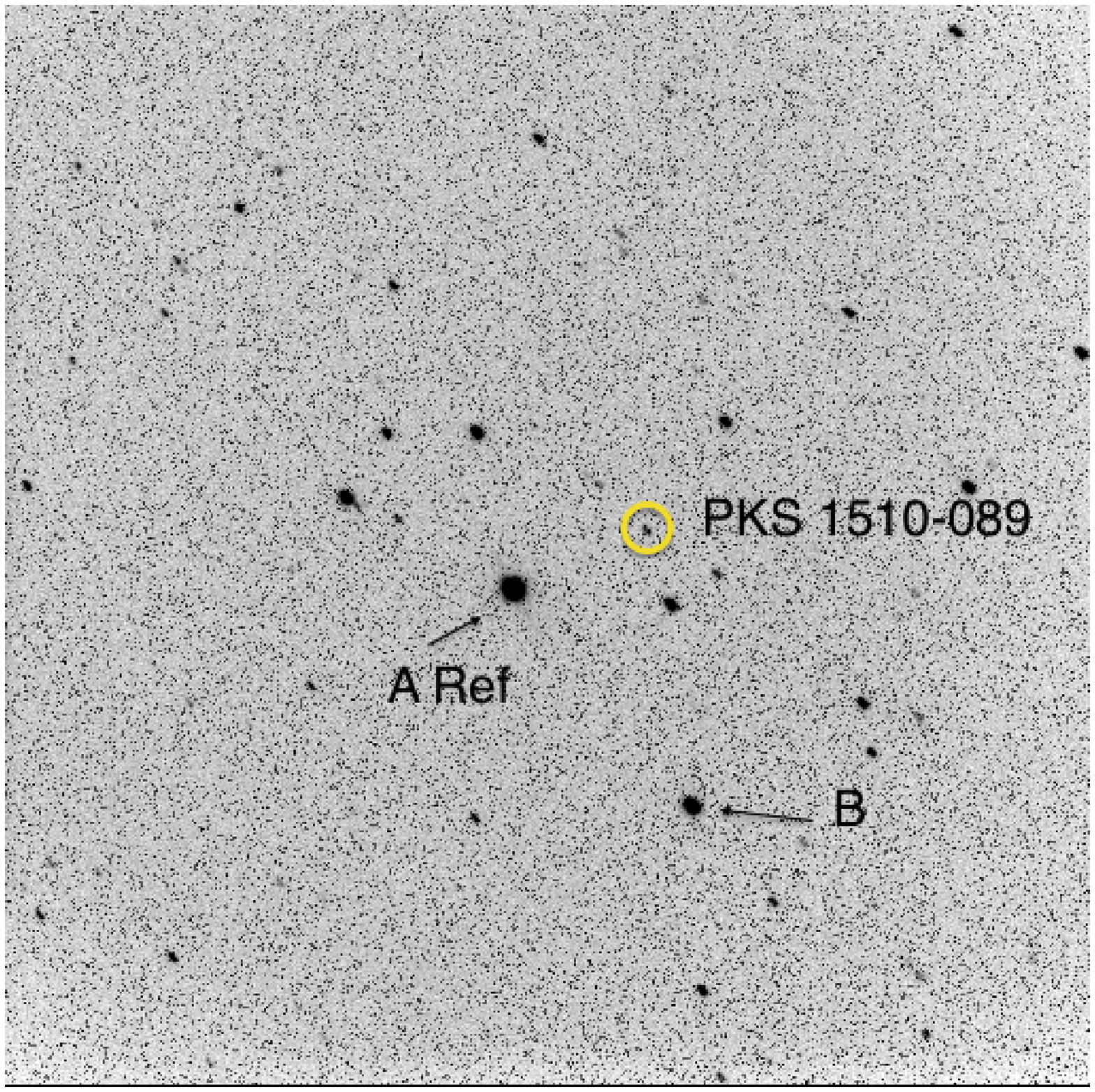}& 
\includegraphics[width=0.6\columnwidth]{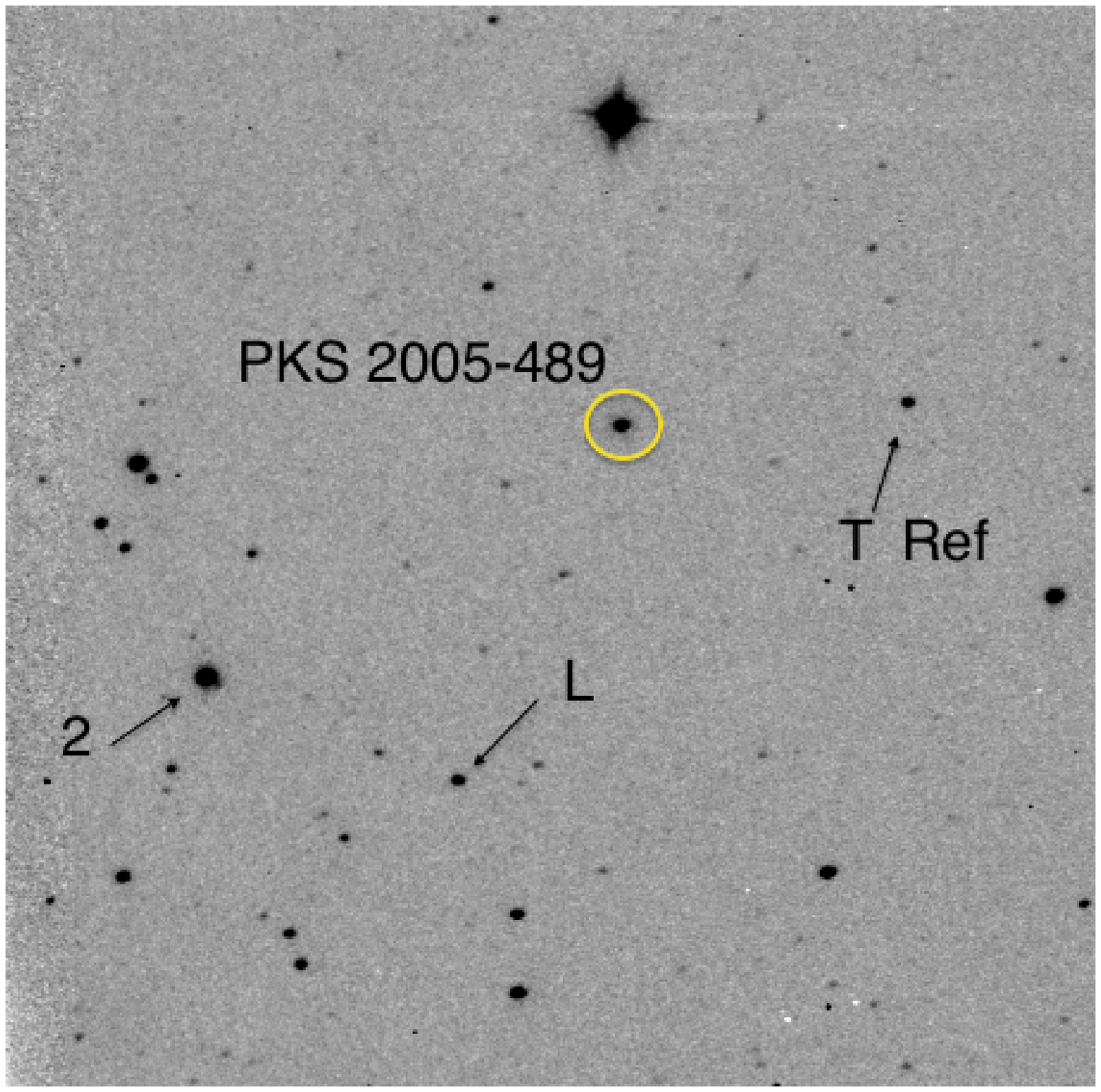}&
\includegraphics[width=0.6\columnwidth]{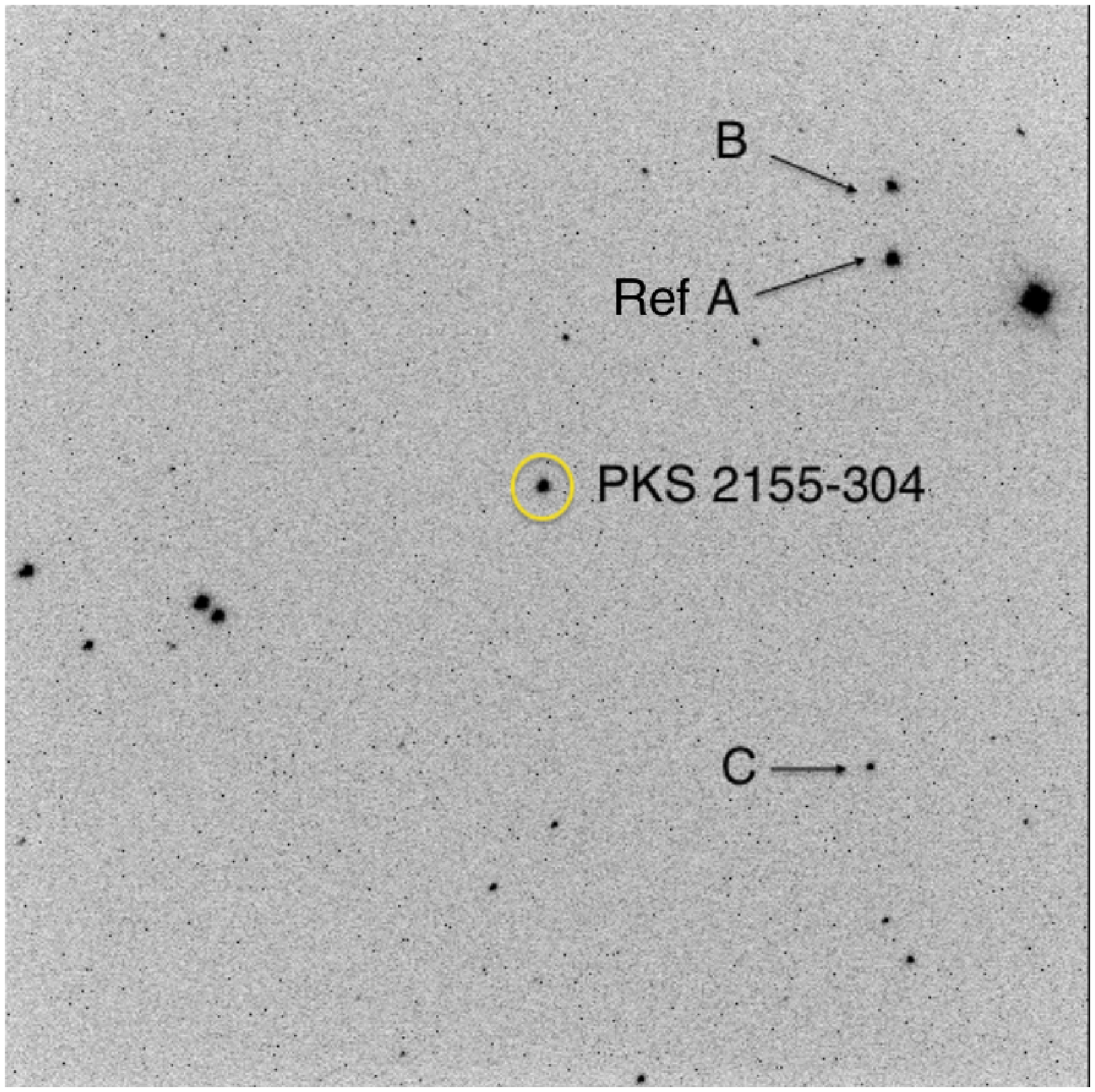}\\
\includegraphics[width=0.6\columnwidth]{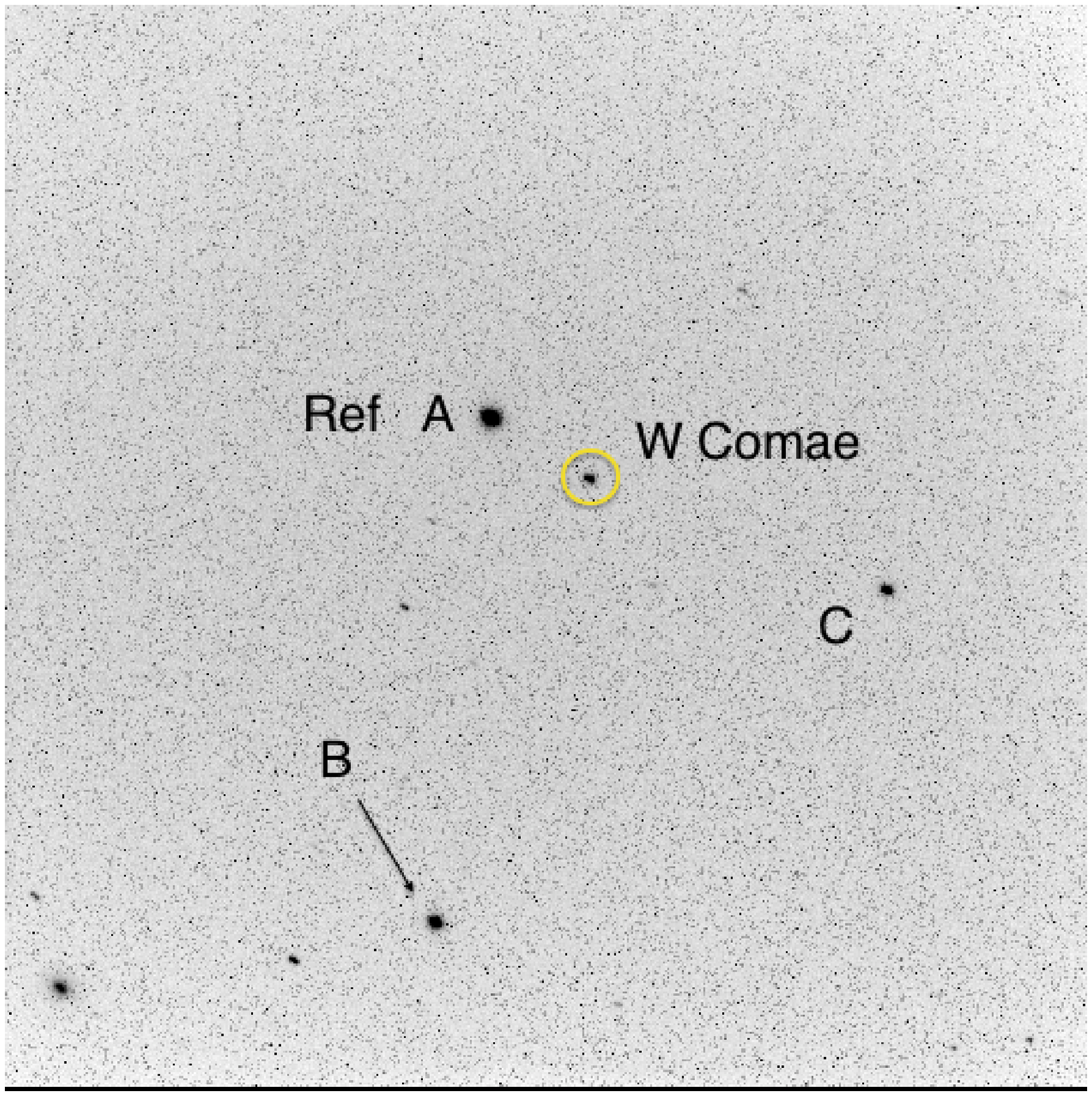}&& \\
\end{tabular}
\caption{Charts for the seven targets in R filter.}
\label{Charts}
\end{figure*}

In Fig.\ref{Charts}  we report some images of the CCD fields where the target, 
the reference, and the check  stars  are indicated. The magnitudes of the 
reference and the check stars are given in Table \ref{ss}   for  V, R, and I 
optical filters. The procedure described in the previous section was
followed.

\begin{table*}
\centering
\small
 \caption{V, R, I Comparison stars in the fields of our sample.}
 \label{ss}
 \begin{tabular}{llccccccccl}
 \hline \hline
  Source		&Star		&Coordinates				&V    		 	&$\sigma$(V)  &	R 		&$\sigma$(R)  &	I 			&$\sigma$(I) &References\\
		        &                 		&R.A.  DEC.       	 		&			&			&			&			&				&			&		\\  
		        &                 		&[h:m:s]   [d:m:s]        			&			&			&			&			&				&			&		\\
\hline
\hline 
&&&&&&&&&\\
PKS 0537-441	&A (Opt Ref)	&	05:39:04.51 -44:06:38.2 	&	10.680	& 	0.010&	 	10.400	&	  0.010	&		10.120 	&  	0.010	&      	1	\\
       			&B			&	05:38:49.33 -44:01:30.0	& 	13.200	&	0.010&	 	12.870	&	  0.010	&	   	12.540	 & 	 0.010 	&     	1	\\
      			&C			&	05:38:59.97 -44:01:19.1	& 	14.080	&	0.010&	 	13.710	&	  0.010	&	   	13.340 	& 	 0.010  	&    	1	\\
    			&E (NIR Ref )	&	05:38:55.70 -44:06:19.6	& 	14.843	&	0.035&	 	14.151	&	  0.034	&	   	13.585	&  	 0.023 	&    	6	\\
&&&&&&&&&\\
\hline
&&&&&&&&&\\
 PKS 0735+17	&C1			&	07:38:00.57 +17:41:19.7	&	13.26	&	 0.04	 &		12.89	&	  0.04	&	   	12.57 	&   	0.04   	&     	2	 \\
 			&C2 (Ref)		&	07:38:08.55 +17:40:29.2	&	13.31	&	 0.04	 &		12.79	&	  0.04	&	   	12.32	&   	 0.04  	&      	2	\\
&&&&&&&&&\\
\hline
 &&&&&&&&&\\
  OJ 287		&O (NIR Ref)	&	08:54:53.38 + 20:04:44.8	&	14.192 	&  	0.003 &	 	13.707	&   0.002 		&   	  	13.262 	&  0.004		&	3	\\
 			&B (Opt Ref)	&	08:54:54.45 + 20:06:13.6	& 	14.627	&  	0.003 &  		14.315	&   0.003 		&   	  	13.999	&   0.004		&	3	\\
  			&E 			&	08:54:55.20 + 20:05:42.3	& 	14.974	&   	0.003 &  		14.632	&   0.003		&  	   	14.304	&   0.004		&	3 	\\
&&&&&&&&&\\
\hline 
&&&&&&&&&\\
PKS 1510-089	&A (Ref)		&	15:12:46.16 -09:05:23.0	&	11.571 	&	 0.001&	 	11.195	&    0.002		&	   	10.847	&   0.001		&	3	\\
			&B 			&	15:12:53.19 -09:03:42.4	& 	13.282	& 	 0.016&	 	12.992 	&   0.022		&	   	12.687	&   0.019		&	3	\\
&&&&&&&&&\\
\hline
&&&&&&&&&\\
PKS 2005-489	&T (Ref)		&	20:09:40:03 -48:50:21.8 	&	14.92	&	 ---&	 		14.39	&	  ---		&   	   	13.86  	&  0.06 		& 	4 (V, R band) ; 6 (I band)    \\
			&L 			&	20:09:19.04 -48:46:43.1 	&	14.41	&	 ---& 			13.89	&  	  ---		&	   	 13.45  	&  0.09		& 	4 (V, R band) ; 6 (I band) 	 \\
    			&2			&	20:09:05.42 -48:47:20.8 	&	11.92	&	 0.03&	     	11.33	&	  0.03	&        	10.79	& 0.03		&     	6 	 \\
&&&&&&&&&\\
\hline
&&&&&&&&&\\
PKS 2155-304	&A (Ref)		&	21:59:02.49 -30:10:46.6&		12.050	&	 0.010&	 	11.670	&	0.010  	&		11.300	 &  0.020		&	1 	\\
			&B 			&	21:59:05.34 -30:10:51.0&		13.000	&	 0.010&	 	12.470	&   0.020  	 	&		12.020 	&  0.020 		&  	1	\\
			&C 			&	21:58:42.33 -30:10:27.4&		14.280	&	 0.010& 		13.920	&   0.020   	&		13.560 	&  0.020		&	1	\\
&&&&&&&&&\\
\hline
&&&&&&&&&\\
W Comae		&A (Ref)		&	12:21:33.66 +28:13:04.1 &	12.08	&	 0.02	& 		11.72	&	   0.04	&	   	11.40 	&   0.04 		&      	5	\\
       			&B 			&	12:21:13.83 +28:13:04.3 &	13.03	&	 0.07&		 12.69	&	   0.05	 &		12.37 	&   0.05   		&	6	\\
			&C 			&	12:21:28.61 +28:16:37.7 &	14.81	&	 0.04	 &		13.86	&	   0.04	&	   	12.68 	&   0.04  		&	5	\\&&&&&&&&&\\
\hline

\end{tabular}
\tablefoot{Stars used as references are marked with Ref;   NIR Ref and  Opt Ref specify that the reference stars change between NIR and optical bands. The check star can  change depending on its positions in the frames.}
\tablebib{
(1) \cite{Hamuy1989};
(2) \cite{Ciprini2007}; (3)  \cite{Gonzalez2001}; (4) \cite{Rector2003};
(5) \cite{Fiorucci1996}; (6)  this work.
}

 \end{table*}

\section{Long term variability}

The light curves resulting from the averaging of the single 
measurements on each night are presented in Fig.\ref{0537lcave} and in Table \ref{photometry}. A full version of  Table \ref{photometry} is available 
in electronic format.
If more data are present in a night interval, weighted averaged magnitudes are taken.  The   standard errors of source and of the check star added in quadrature with the mean instrumental and calibration error is assumed as uncertainty. 
For each source the corresponding check star light curve is reported in the J filter and  in the other bands the curve for the check star is very similar.

 \setcounter{figure}{1}
\begin{figure*}
\centering
\includegraphics[width=1.8\columnwidth]{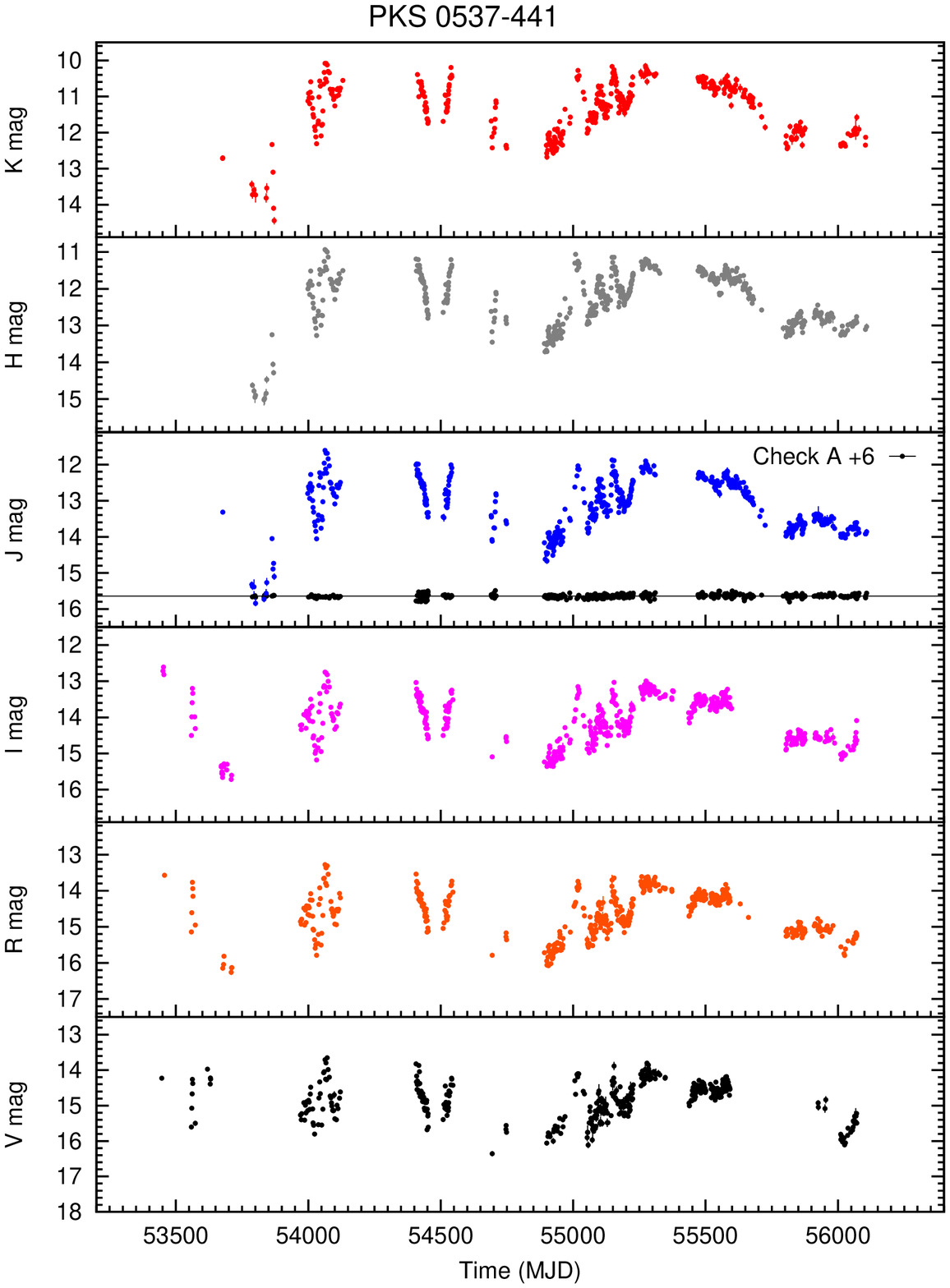}
\caption{REM near-infrared and optical nightly averaged light curves of the seven target sources. The light curve of the check star is reported in J band (black points) with the indicated displacements $\Delta$m.}
\label{0537lcave}
\end{figure*}

 \setcounter{figure}{1}   
\begin{figure*}
\centering
\includegraphics[width=1.8\columnwidth]{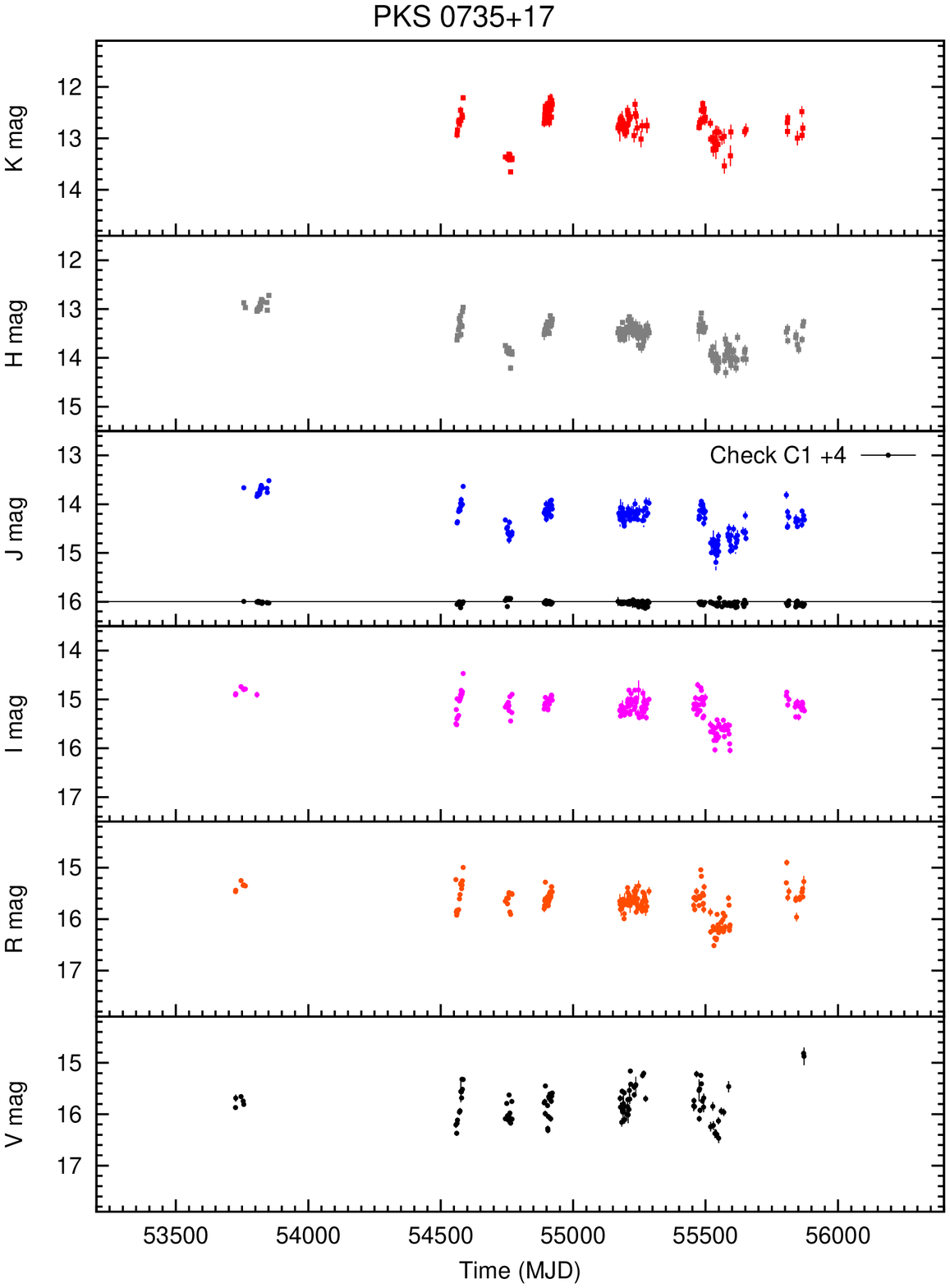}
\caption{--- Continued}
\label{0735lcave}
\end{figure*}

 \setcounter{figure}{1}   
          \begin{figure*}
           \centering
   \includegraphics[width=1.8\columnwidth]{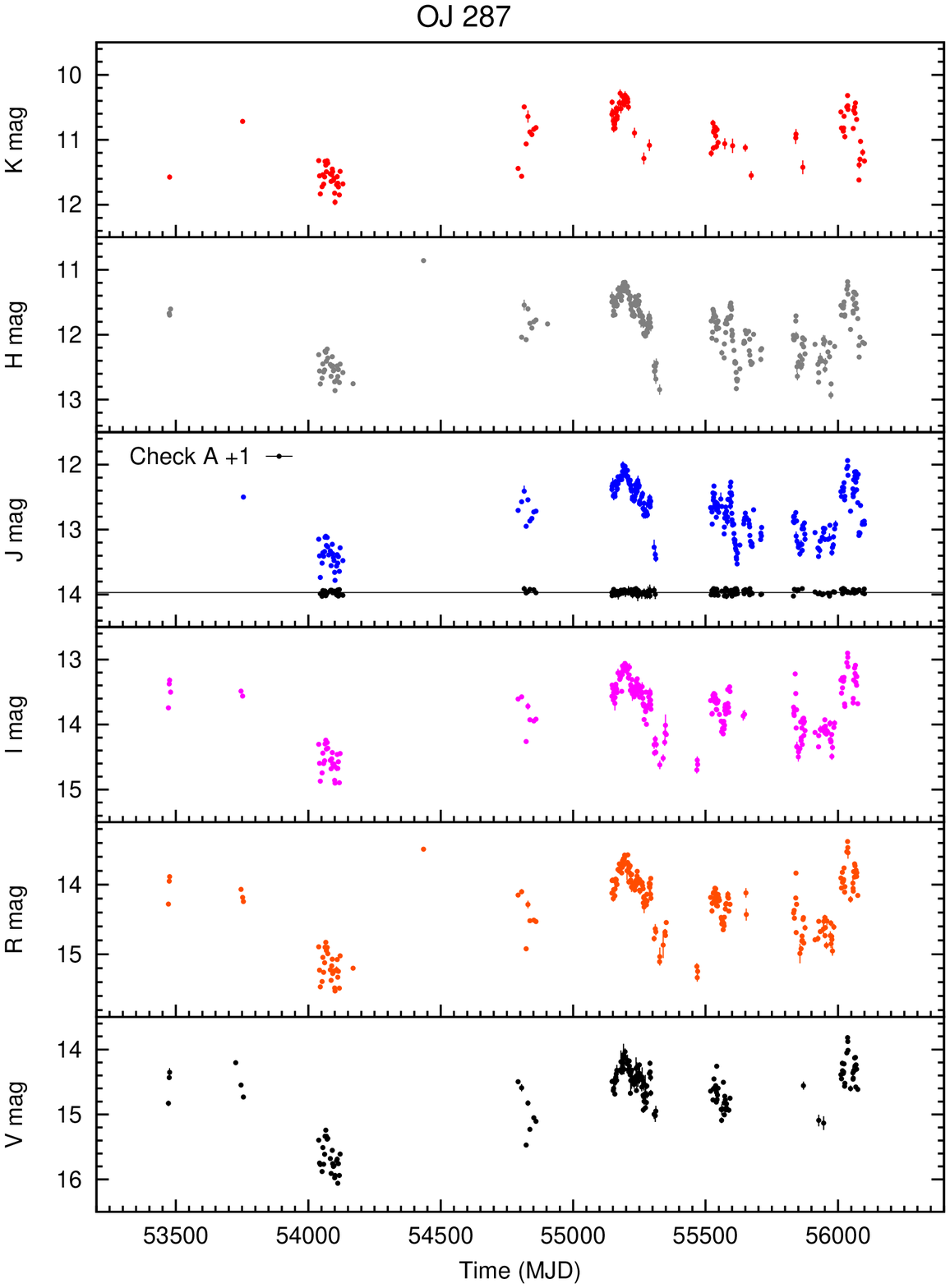}
      \caption{--- Continued}
         \label{OJlcave}
   \end{figure*}

 \setcounter{figure}{1}        
       \begin{figure*}
          \centering
   \includegraphics[width=1.8\columnwidth]{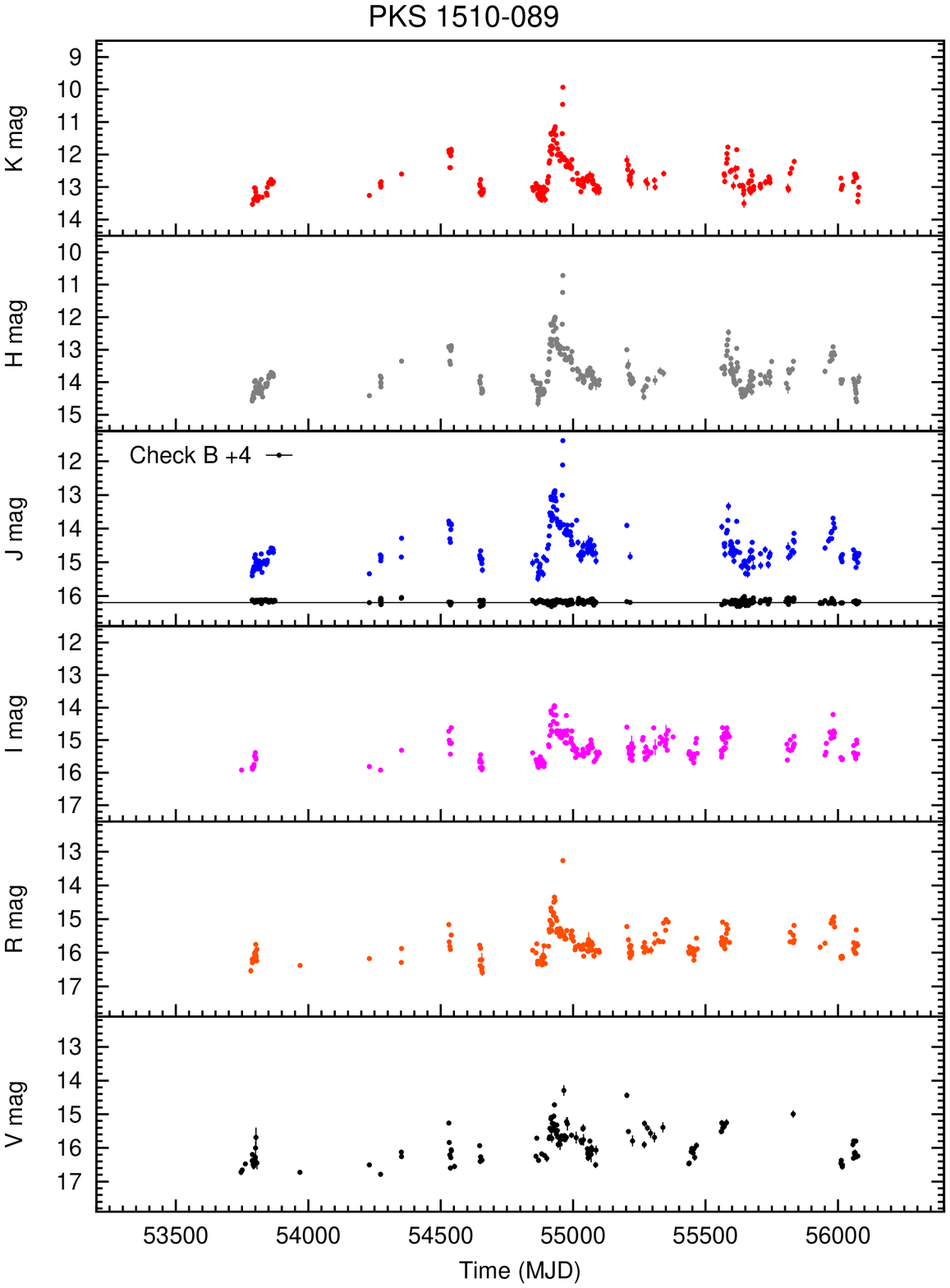}
      \caption{--- Continued}
         \label{1510lcave}
   \end{figure*}

  \setcounter{figure}{1} 
     \begin{figure*}
        \centering
   \includegraphics[width=1.8\columnwidth]{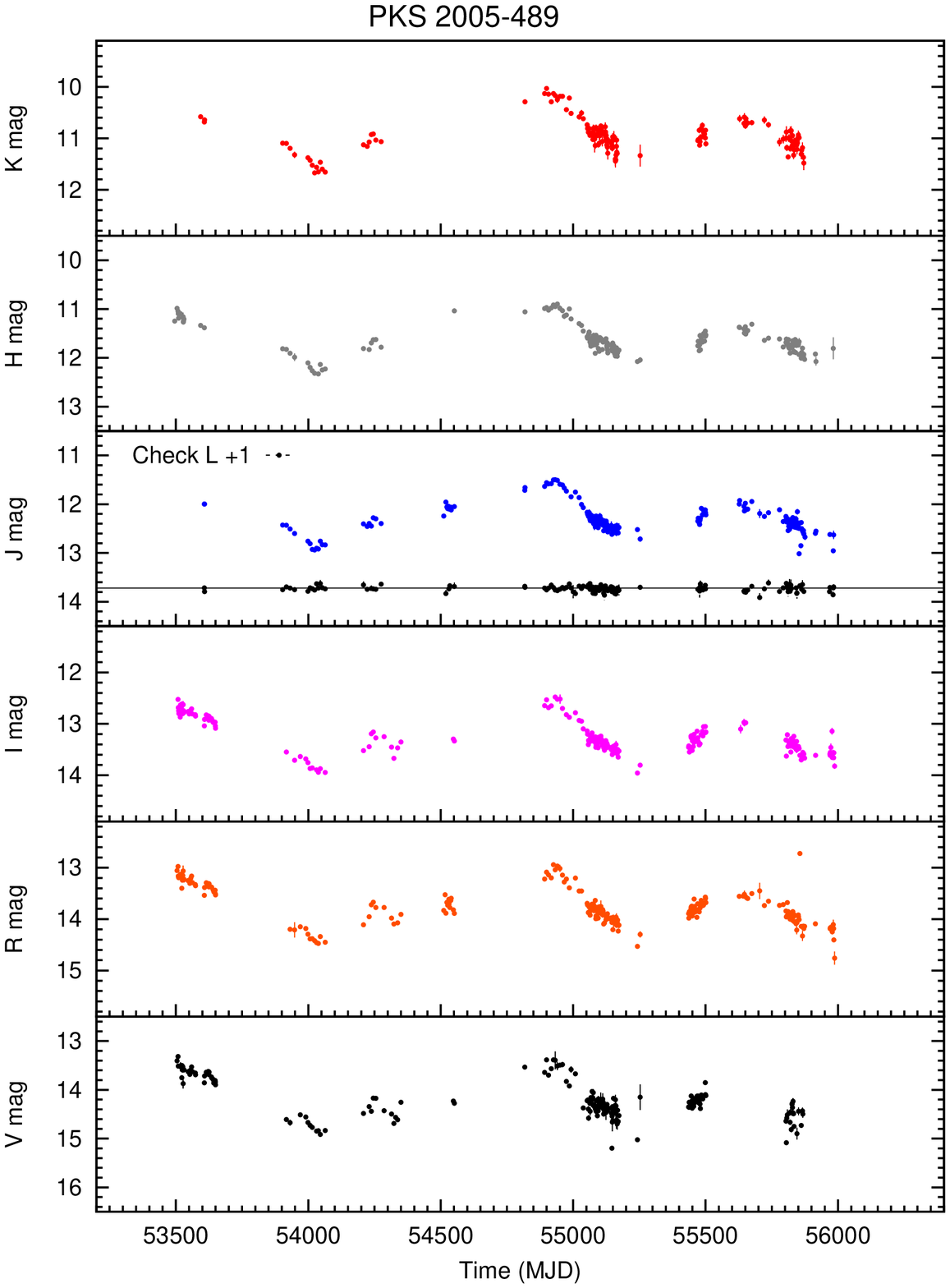}
      \caption{--- Continued}
         \label{2005lcave}
   \end{figure*}

 \setcounter{figure}{1}
      \begin{figure*}
        \centering
   \includegraphics[width=1.8\columnwidth]{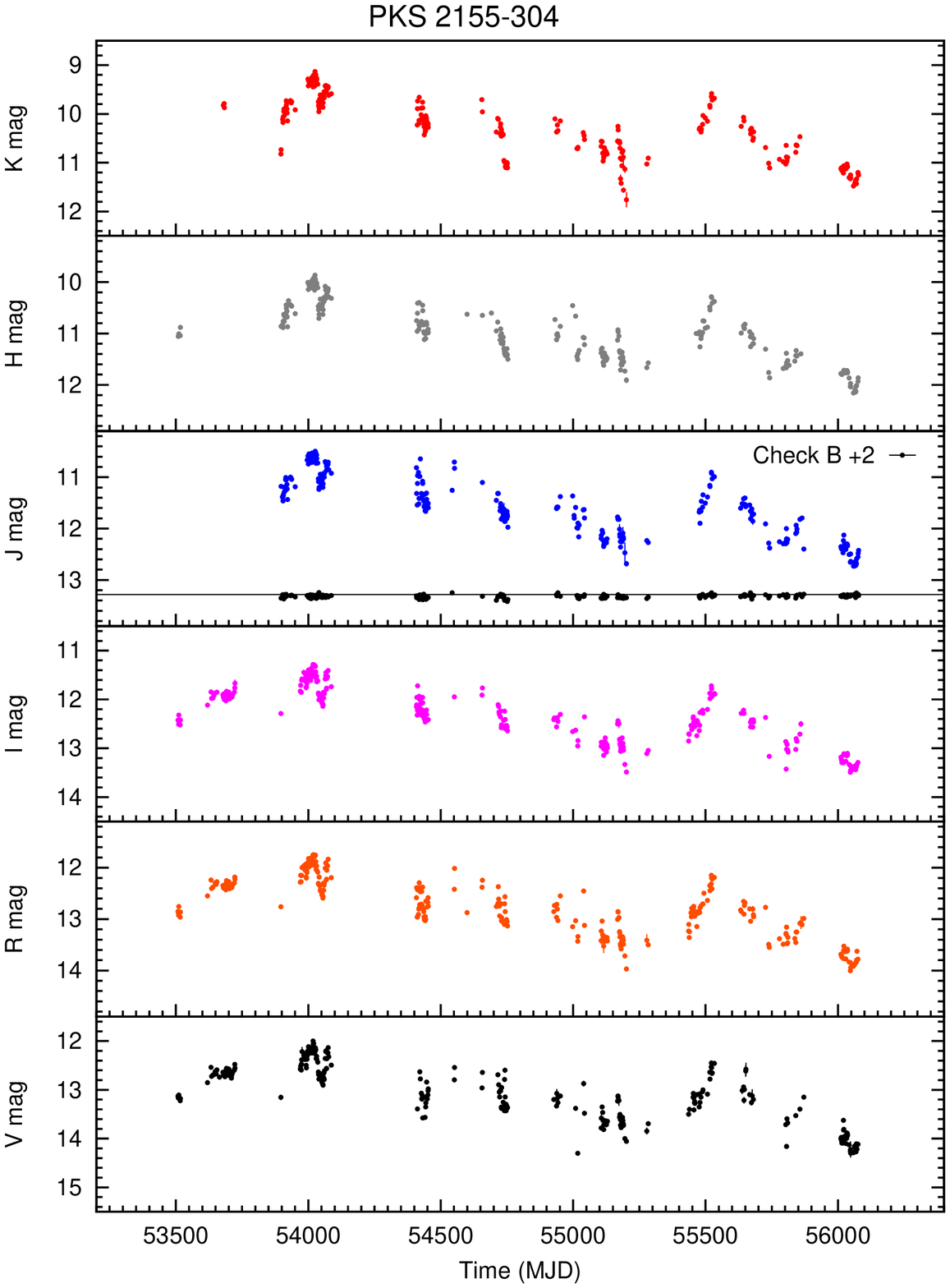}
      \caption{--- Continued}
         \label{2155lcave}
   \end{figure*}

   \setcounter{figure}{1}
          \begin{figure*}
          \centering
   \includegraphics[width=1.8\columnwidth]{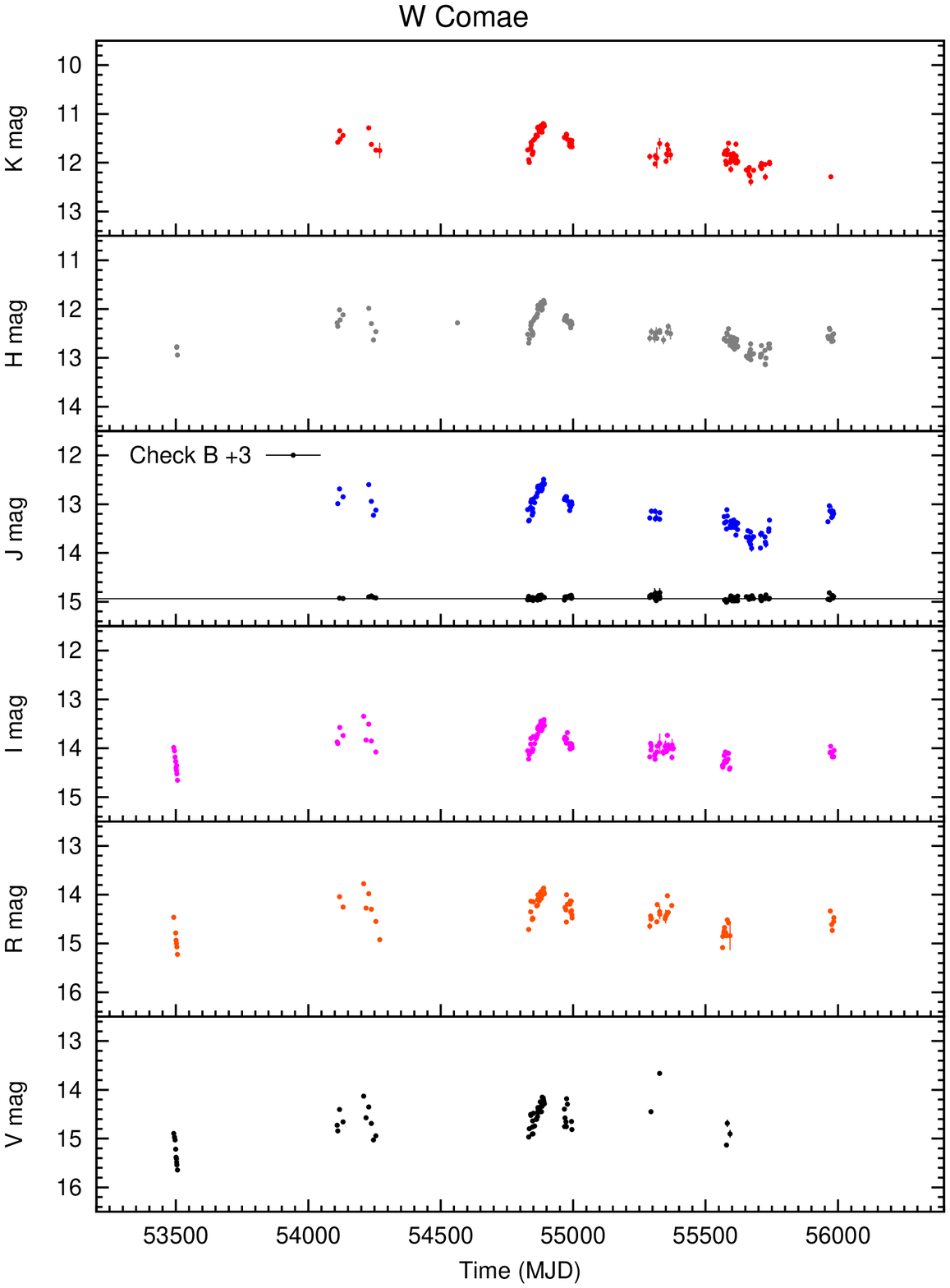}
      \caption{--- Continued}
         \label{WClcave}
   \end{figure*}

\begin{table*}
\centering
\caption{Photometry of the seven target objects.} 
\begin{tabular}{ccccc}
\hline\hline
	Source		&Filter	&Time of			&Average 	& Magnitude \\	
				&		&observation		&magnitude	&error		\\
				&		&[MJD]			&[mag]		&	[mag]\\
\hline\hline
&&&&\\	

PKS0537-441	&	K		&	53453 		&  	10.685  		&0.031\\
PKS0537-441	&K			&	53676		&  	12.716  		&0.044 \\
PKS0537-441	&K			&	53677		&  	12.699 		&0.049 \\
	....		&....			&  	....			&	....			&  	.... \\
&&&&\\
\hline
\label{photometry}
\end{tabular}		
\end{table*}

\begin{table*}
\centering
\scriptsize
\caption{Properties of the light curves.}
\begin{tabular}{lcccccccccc}
\hline\hline             

	    Source	&       Timeframe			&Filter	&	Magnitude  	&Variability		&   Average	&  	Mode  	&	Median	&	STDev.	&Flux Range		&$F_{max}/F_{min}$	\\ 
                         	&        in H band			&         	& 	range	  	&amplitude		&    magnitude  	&      			&			& 		 	&				&	\\
                          	&	([MJD])				&	 	&	 [mag]		&$\Delta$m [mag] 	&	[mag]	&	  [mag]	&	[mag]	&	[mag]	&[mJy]			&	\\
\hline 
\hline
&&&&&&&&&&\\
PKS 0537-441	&  2005/11/03 - 2012/06/30	& K 		& 10.06 - 14.57 	& 4.51 			& 11.30 		& 10.6 		& 11.15 		& 0.74 		& 1.14 - 63.07 		& 55.32\\
 			& 	(53677 - 56108)		& H 		& 10.93 - 15.45 	& 4.52 			& 12.28 		& 11.8 		& 12.18 		& 0.73 		& 0.70 - 45.16 		& 64.51\\
			& 						& J 		& 11.58 - 15.84 	& 4.26 			& 13.09 		& 12.5 		& 13.01 		& 0.71 		& 0.74 - 37.41		& 50.55\\
			&  						& I 		& 12.59 - 15.75 	& 3.16 			& 14.08		 & 13.6 		& 14.06		 & 0.63 		& 1.29 - 23.79 		& 18.44\\
 			& 						& R 		& 13.27 - 16.28 	& 3.01 			& 14.65 		& 14.2 		& 14.65 		& 0.61 		& 1.04 - 16.64 		& 16.00\\			 
			&						& V 		& 13.11 - 16.36 	& 3.25 			& 14.83 		& 14.6 		& 14.78		 & 0.53 		& 1.18 - 23.49 		& 19.91\\			 
 &&&&&&&&&&\\

PKS 0735+17	& 2006/01/19 - 2011/11/05 	& K 		& 12.14 - 13.65 	& 1.51 			& 12.78 		& 12.7 		& 12.71		& 0.32 		& 2.35 - 9.41		& 4.00\\ 
 			& 	(53754 - 55870) 		& H 		& 12.72 - 14.30 	& 1.58 			& 13.52 		& 13.4 		& 13.46 		& 0.34 		& 2.02 - 8.65		& 4.28\\
 			&  						& J 		& 13.52 - 15.19 	& 1.67 			& 14.28		& 14.2 		& 14.23 		& 0.34 		& 1.34 - 6.27 		& 4.68\\
			& 						& I 		& 14.47 - 16.04 	& 1.57 			& 15.19 		& 15.1 		& 15.14 		& 0.28 		& 0.98 - 4.20	 	& 4.29\\
 			& 	 					& R 		& 14.90 - 16.52 	& 1.62 			& 15.68 		& 15.6 		& 15.64 		& 0.28		& 0.83 - 3.68 		& 4.43\\
			&						& V 		& 14.72 - 16.46 	& 1.74 			& 15.79 		& 15.7 		& 15.79		& 0.32 		& 1.05 - 5.24 		& 4.99\\
&&&&&&&&&&\\

OJ 287		& 2005/04/11 - 2012/06/21 	& K 		& 10.27 - 11.96 	& 1.69 			& 10.99 		& 10.4 		& 10.88		 & 0.46 		&11.12 - 52.94 		& 4.76\\ 
 			& (53471 - 56099) 			& H 		& 10.84 - 12.93 	& 2.09 			& 11.92 		& 11.7 		& 11.84 		& 0.44 		& 7.10 - 48.82 		& 6.88\\
 			& 						& J 		& 11.94 - 13.78 	& 1.84			& 12.74 		& 12.3 		& 12.69 		& 0.42 		& 4.90 - 26.76 		& 5.46\\
			&  						& I 		& 12.90 - 14.90 	& 2.00 			& 13.79 		& 13.5 		& 13.70 		& 0.45 		& 2.80 - 17.64 		& 6.30\\
			& 						& R 		& 13.38 - 15.52 	& 2.14 			& 14.31 		& 14.0 		& 14.19 		& 0.48 		& 2.05 - 14.74 		& 7.19\\
			&						& V 		& 13.82 - 16.06 	& 2.24 			& 14.69 		& 14.4 		& 14.55 		& 0.49 		& 1.51 - 11.91 		& 7.89\\
&&&&&&&&&&\\
 
PKS 1510-089	& 2006/01/27 - 2012/06/01 	& K 		& 9.93 - 13.78 		& 3.85 			& 12.67 		& 12.8 		& 12.82 		& 0.56 		& 2.14 - 73.8		 & 34.49\\
 			& (53783 - 56079)			& H 		& 10.72 - 14.80 	& 4.08 			& 13.74 		& 14.0 		& 13.89 		& 0.59 		& 1.31 - 56.58 		& 43.19\\
 			& 						& J 		& 11.39 - 15.57 	& 4.18 			& 14.54 		& 14.7 		& 14.69 		& 0.61 		& 1.00 - 47.17 		& 47.17\\
 			&  						& I 		& 13.95 - 15.97 	& 2.02 			& 15.21 		& 15.4 		& 15.28 		& 0.40 		& 1.18 - \ 7.62 		& 6.46\\
 			&  						& R 		& 13.26 - 16.69 	& 3.43 			& 15.73 		& 15.9 		& 15.79 		& 0.44 		& 0.83 - 19.44 		& 23.42\\
			&						& V 		& 14.29 - 16.93 	& 2.64 			& 15.90 		& 16.2 		& 15.94 		& 0.50 		& 0.83 - \ 9.44 		& 11.37\\
&&&&&&&&&&\\ 

PKS 2005-489	& 2005/04/20 - 2012/02/25 	& K 		& 10.03 - 11.67 	& 1.64 			& 10.96		 & 11.1 		& 10.98 		& 0.32 		& 14.61 - 66.42 	& 4.55\\
 			&  (53480 - 55982)			& H 		& 10.88 - 12.33 	& 1.45 			& 11.64 		& 11.7 		& 11.69 		& 0.29 		& 12.55 - 47.53 	& 3.79\\
 			& 						& J 		& 11.45 - 13.01 	& 1.56 			& 12.32 		& 12.4 		& 12.37 		& 0.28 		& 10.17 - 42.85		& 4.21\\
 			&  						& I 		& 12.09 - 13.96 	& 1.87 			& 13.27 		& 13.3 		& 13.34 		& 0.31 		& \ 6.99 - 38.97 	& 5.58\\
 			&  						& R 		& 12.72 - 14.76 	& 2.04 			& 13.76		 & 13.9 		& 13.83 		& 0.34 		& \ 4.42 - 28.81 	& 6.52\\
			&						& V 		& 12.98 - 15.20 	& 2.22 			& 14.19 		& 14.2 		& 14.27 		& 0.39 		& \ 3.61 - 27.92 	& 7.73\\
&&&&&&&&&&\\ 

PKS 2155-304	& 2005/05/18 - 2012/05/29 	& K 		& 9.08 - 11.57 		& 2.49 			& 10.03 		& 9.3 		& 10.55 		& 0.62 		& 15.92 - 157.49 	& 9.89\\
 			& (53508 - 56076) 			& H 		& 9.81 - 12.16 		& 2.35 			& 10.94		 & 11.6 		& 11.43 		& 0.62 		& 14.40 - 124.96 	& 8.68\\
 			& 						& J 		& 10.44 - 12.75 	& 2.32 			& 11.40 		& 10.6 		& 11.35 		& 0.63 		& 12.57 - 106.07 	& 8.44\\
 			& 						& I 		& 11.17 - 13.53 	& 2.36 			& 11.92 		& 11.9 		& 11.90 		& 0.37 		& 9.78 - 85.69 		& 8.77\\
 			&  						& R 		& 11.63 - 14.04 	& 2.41 			& 12.40 		& 12.3 		& 12.32		& 0.41 		& 7.88 - 72.53 		& 9.20\\
			&						& V 		& 11.98 - 14.30		& 2.32			& 12.68		 & 12.6 		& 12.63		& 0.39 		& 7.49 - 64.43 		& 8.47\\
&&&&&&&&&&\\

W Comae 	&         2005/05/12 - 2012/02/26&K		&   	11.20 - 12.39	&         1.19		&	11.71	& 	11.3		&      11.72	&	0.30		&	7.47 - 22.42	&	3.00	\\	
			&		(53502 - 55983)	&H		&   	11.83 - 13.14	&         1.31		&      	12.48	&	12.7		&	12.53	&	0.33		& 	5.84 - 19.60	&	3.36\\
		 	&                  				&  J		& 	12.49 - 14.15	&        1.66			&        13.17	&	13.4		&	13.17	&	0.37		& 	3.48 -16.07	&	4.62\\
			&                  		 		&  I		& 	13.35 - 14.67	&        1.33			&         13.93	&	14.0		&	13.97	&	0.28		& 	3.42 - 11.61	&	3.39\\
			&						&  R		& 	13.78 - 15.24	&         1.46		&         14.36	&	14.0		&	14.34	&	0.33		& 	2.64 - 10.11	&	3.83\\								&						&  V		& 	13.97 -15.66	&         1.69		&         14.65	&	14.4		&	14.64	&	0.39		&	2.16 - 10.22	& 	4.73 \\	
&&&&&&&&&&\\		 
\hline
\end{tabular}
\label{lc}
\end{table*}

\begin{table*}
\centering
\small
\caption{Absorption coefficients}
\begin{tabular}{ccccccccc}
\hline\hline
			&$\lambda$	&PKS 0537-441	&PKS 0735+17 	& OJ 287		&PKS 1510-089	&PKS 2005-489	&PKS 2155-304	&W Comae 	\\	
			&[$\mu$m]	&				&				&			&				&				&				&\\
\hline\hline
&&&&&&&&\\	
A$_{\rm K}$	&	2.15		&	0.0137		&  	0.0124  		&0.0103		&0.0369			&0.0205			&0.0080		&0.0085	 	\\
A$_H$		&  	1.64		&	0.0212		&  	0.0192  		&0.0160 		&0.0571			&0.0317			&0.0124 		&0.0132		\\
A$_J	$		&  	1.25		&	0.0328		&  	0.0297 		&0.0248		&0.0883			& 0.0491			&0.0192		&0.0204		 \\
A$_I	$		&  	0.804	&	0.0685		&  	0.0620 		& 0.0517		&0.1845			&0.1026			&0.0400		&0.0426		\\
A$_R$		&  	0.659	&	0.0946		&  	0.0856 		&0.0714		&0.2548			& 0.1417			&0.0552		&0.0589		\\
A$_V$		&  	0.55		&	0.1162		&  	0.1051		&0.0878		&0.3128			&0.1740			&0.0678		&0.0723		\\
&&&&&&&&\\
\hline
\label{abs}
\end{tabular}		
\tablefoot{Extinction values are derived from the NASA/IRPAC Infrared Science Archive \citep{Schlegel1998}.  \cite{Cardelli1989} formulae were used  to calculate\footnote{http:// dogwood.physics.mcmaster.ca/Acurve.html} the absorption coefficients, assuming the extinction to reddening ratio Av / E(B-V) = 3.1.}
\end{table*}

\begin{figure}
\centering
\includegraphics[angle=-90,width=9cm]{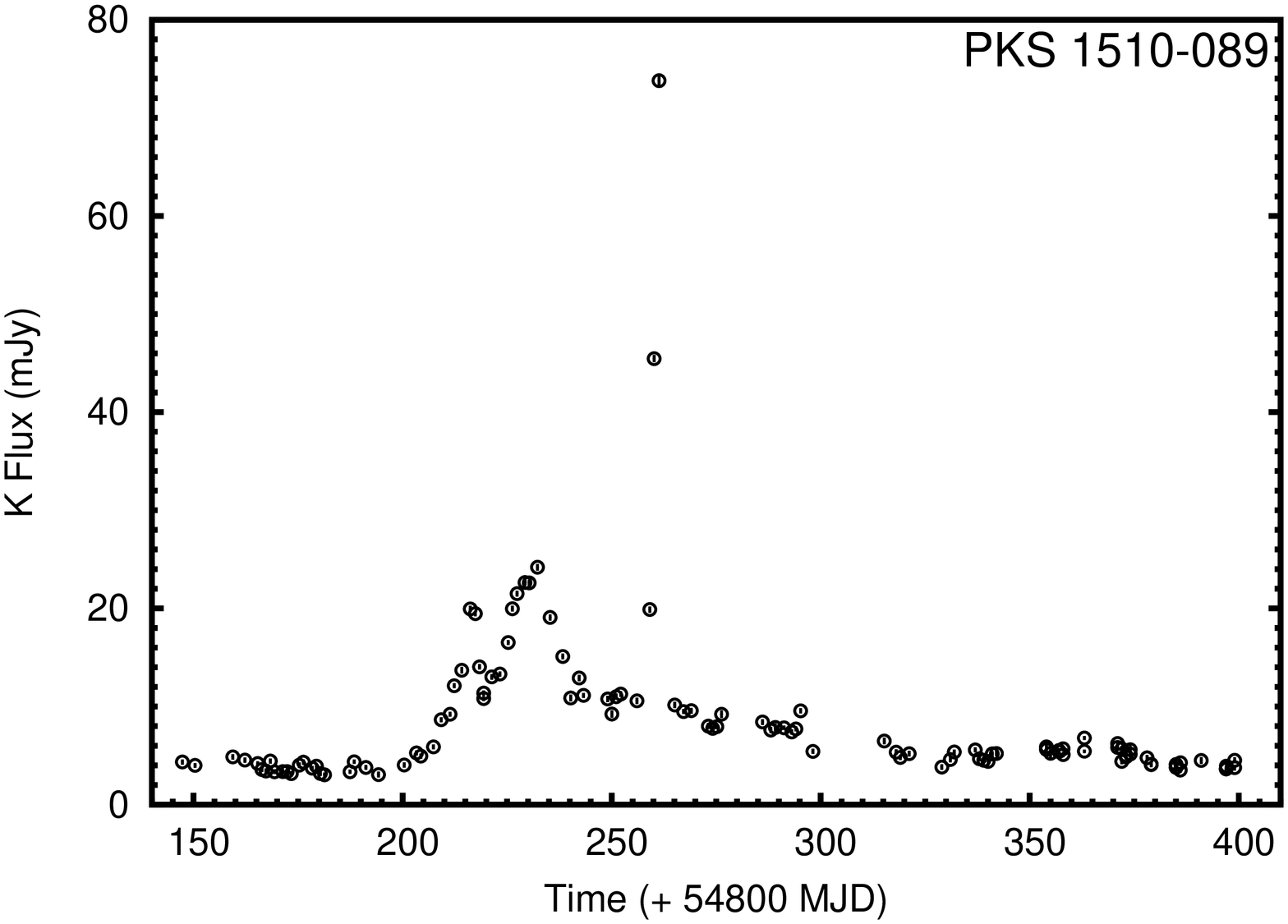}
\caption{Active state of PKS 1510-089 and the flare occurring around 
54961 MJD.}
\label{1510flare}
\end{figure}

  The properties of the light curves are reported in Table \ref{lc}, where the flux values are given after dereddening with the coefficient reported in Table \ref{abs}. All seven sources are strongly variable.  The amplitude of variability  is larger than four magnitudes in the NIR bands of PKS 0537-441and PKS 1510-089, decreasing in the optical.  For the BL Lacs PKS 0735+17 , OJ 287,  PKS 2005-489, and W Comae the variability is 1-2 mag, while the case of PKS 2155-304 is intermediate.  The curves  are spiky, with monotonic trends that can last for months. There are flares of various intensities and shapes, and their classification appears arduous. As an example we select in Fig.\ref{1510flare}   
  the prominent flare that occurred in PKS 1510-089  and lasted for nine days  around May 10, 2009 (54961 MJD) while the source was brightening for 120 days. The source K flux dramatically rose from $\sim$10 to 73.8 mJy in about six days before decaying in four days to the original value. The noticeable event was also observed simultaneously by \cite{Sasada2011} and \cite{Bonning2012}.
  
  We also characterized the variability of all the light curves  through the fractional variability amplitude  $\sigma_{rms}$ defined as
  
\begin{equation}
\sigma_{rms}^{2}=\frac{1}{N\mu^{2}}\sum\left[\left(F_{i}-\mu\right)^{2}-\sigma_{i}^{2}\right] ,
\label{EV}
\end{equation}
where  $N$ is the number of flux values $F_{i}$, with measurement  uncertainties $\sigma_{i}$, and  $\mu$ is the average flux. The evaluation of $\sigma_{rms}$ gives a measure of the intrinsic variability amplitude
and it represents  the averaged amplitude of observed variations as a percentage of the flux corrected for the effects of measurement noise. It is discussed, for example,  in  \cite{Nandra1997}, \cite{Edelson2002},  and \cite{Vaughan2003}.  
From Table \ref{EVave}  and Fig.\ref{EV} it is apparent that in terms of $\sigma_{rms}$ the most variable source is  PKS 1510-089, followed by PKS 0537-441 and PKS 2155-304,  consistently with the variability indicated by the secular excursion of the source magnitude (see above). The dependence of $\sigma_{rms}$  on the spectral band  is modest in the BL Lacs objects and is possibly  in part due to the dishomogeneity in the coverage. The FSRQ PKS 1510-048 is more variable at longer wavelengths. The latter behavior with smoother trends is also pointed out by \cite{Bonning2012}.

\begin{table*}         
\centering
\caption{NIR-optical fractional variability amplitude $\sigma_{rms}$ of the blazar sample (x 100 \%)}
\begin{tabular}{llcccccc}     
\hline\hline       
Source 			& Class		& K 				& 	H 			& J 				& I 				& R 				& V\\ 
\hline\hline                    
&&&&&&&\\
PKS 0537-441		& BL Lac 	& 	54 $\pm$ 2 		& 57 $\pm$ 2 	&57 $\pm$ 2	&55 $\pm$ 2 	&52 $\pm$ 2 	&47 $\pm$ 2 \\
PKS 0735+17		& BL Lac 	&	25 $\pm$ 2 		&30 $\pm$ 2 	&29 $\pm$ 2 	&23 $\pm$ 1	&25 $\pm$ 1	&31 $\pm$ 2\\ 
OJ 287			& BL Lac	& 	39 $\pm$ 3 		&38 $\pm$ 2 	&37 $\pm$ 2 	&37 $\pm$ 2	&39 $\pm$ 2	&36 $\pm$ 2\\
PKS 1510-089 	(a)	& FSRQ 	& 	92 $\pm $ 4		&104 $\pm$ 4 	&115 $\pm$ 5 	& 43 $\pm$ 2 	&69 $\pm$ 4 	&52 $\pm$ 4\\
PKS 1510-089 (b)	& 		& 	61 $\pm $ 3		&69 $\pm$ 3 	&68 $\pm$ 3 	& 44 $\pm$ 2 	&45 $\pm$ 2 	&47 $\pm$ 4\\
PKS 2005-489		& BL Lac 	&	32 $\pm$ 2		&29 $\pm$ 2 	&29 $\pm$ 1 	&31 $\pm$ 1 	&33 $\pm$ 1 	&38 $\pm$ 2\\
PKS 2155-30	 	&BL Lac 	& 	57 $\pm$ 3 		&54 $\pm$ 2	&54 $\pm$ 2 	&48 $\pm$ 2 	&49 $\pm$ 2 	&49 $\pm$ 2 \\
W Comae  		& BL Lac 	& 	26 $\pm$ 2 		&31 $\pm$ 2 	& 33 $\pm$ 2	& 25 $\pm$ 2 	&28 $\pm$ 2 	&34 $\pm$ 3\\
&&&&&&&\\
\hline
\end{tabular} 
\label{EVave}   
\tablefoot{

(a): All data;  (b): Because of  inhomogenous sampling, the flare occurring on 54956 - 54969 MJD was removed.}

\end{table*}

      \begin{figure}
        \centering
  \includegraphics[angle=-90,width=8cm]{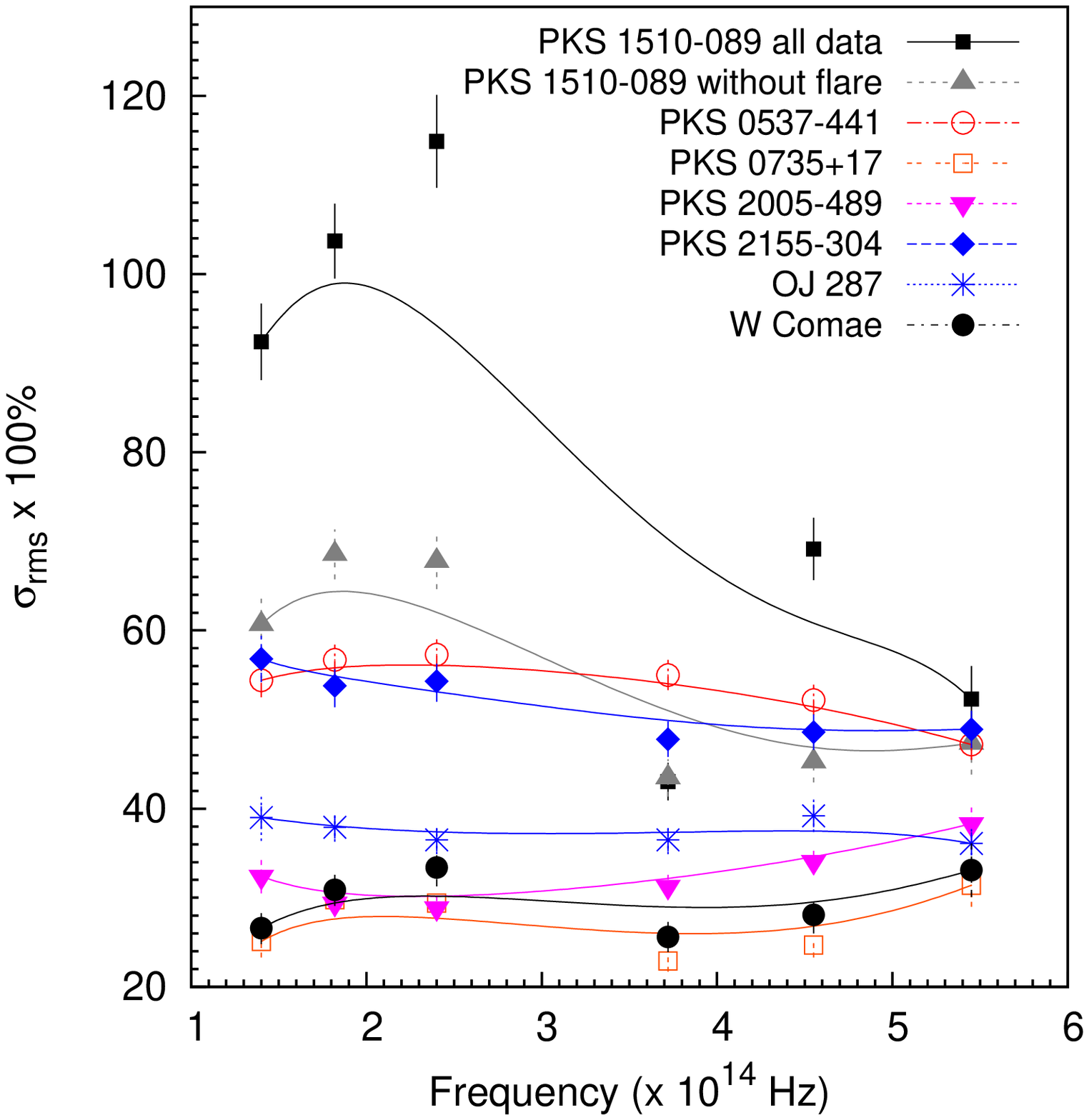}
    \caption{Fractional variability amplitude \textit{vs} frequency. The lines are a help to the eye.}
      \label{EV}
   \end{figure}

   The dependence of $\sigma_{rms}$ on flux was also investigated. The H and R light curves were divided into four flux intervals containing approximatively the same number of points. Average fluxes and $\sigma_{rms}$ are evaluated for each interval, and no clear trend $\sigma_{rms}$ \textit{vs} $F$ is apparent.  The results are collected in Table \ref{EVvsFtab}.  We note that our results do not agree with the finding by \cite{Edelson2013} of a clear correlation between rms and flux  of the BL Lac object W2R1926+42 observed by Kepler satellite.

\begin{table*}  [b]         
\centering
\small
\caption{NIR-optical fractional variability amplitude $\sigma_{rms}$ dependence on flux.}
\begin{tabular}{ccc|ccc|ccc}     
\hline\hline  
\multicolumn{3}{c|}{PKS 0537-441}							&\multicolumn{3}{|c|}{PKS 1510-089}							&\multicolumn{3}{|c}{PKS 2155-304}  \\   		
	Filter	& Average flux				& $\sigma_{rms}$		&Filter	&Average flux			& $\sigma_{rms}$	&Filter	& Average flux			& $\sigma_{rms}$\\ 
		&[mJy]					&(x 100 \%)			&		&[mJy]				&(x 100 \%)		&		&[mJy]				&(x 100 \%)\\
\hline\hline                  

&&&&&&&&\\
	H	&\  \  \  \ 5.7		$\pm$	0.13		&27.5	$\pm$	1.7	&H		&2.08	$\pm$ 0.02 	&7.9 	$\pm$ 1.1		&H		&22.6	$\pm$ 0.5		& 17.3	$\pm$ 1.6\\				
		&10.6	$\pm$	0.2		&19.0	$\pm$	1.2	&		&2.74	$\pm$ 0.02 	& --- 				&		& 35.7	$\pm$ 0.6		& 13.7	$\pm$ 1.2\\
		& 18.4	$\pm$	0.2		&12.3	$\pm$	0.8	&		&3.53	$\pm$ 0.04 	&6.3 		$\pm$1.0	&		& 54		$\pm$ 1		& 15.6	$\pm$ 1.4\\
		&27.1	$\pm$	0.3		&15.1	$\pm$	0.8	&		& 8.8		$\pm$ 0.8		&81		$\pm$ 7	&		&91		$\pm$ 2 		& 16.5	$\pm$ 1.5\\
					
&&&&&&&&\\	
			
	R	&2.31	$\pm$	0.05		&23.8	$\pm$	1.5	&R		&1.31 	$\pm$ 0.02	&10.2 	$\pm$   0.3&	R	&13.0	$\pm$ 0.3		&17.0		$\pm$ 1.4\\	
		&3.80	$\pm$	0.04		&11.5	$\pm$	0.9	&		&1.73	$\pm$ 0.01 	& ---				&		&21.7	$\pm$ 0.2		& 9.4			$\pm$ 0.8\\				
		&6.14	$\pm$	0.06		&11.1	$\pm$	0.8	&		&2.14	$\pm$ 0.02	&5.9 		$\pm$ 1.0	&		&31.8	$\pm$ 0.5		&13.7 		$\pm$ 1.1\\
		&9.3		$\pm$	0.2		&19.0	$\pm$	1.2	&		&3.9		$\pm$ 0.4		& 64 		$\pm$ 7	&		&48.4	$\pm$ 0.9 	&16.0 		$\pm$ 1.3\\		
&&&&&&&&\\
\hline
\end{tabular}
\tablefoot{No evaluation is reported when $\sigma_{rms}^{2}$ is dominated by errors, resulting in a negative value. }  
\label{EVvsFtab} 
\end{table*}

\begin{figure}
\centering
\includegraphics[width=1\columnwidth]{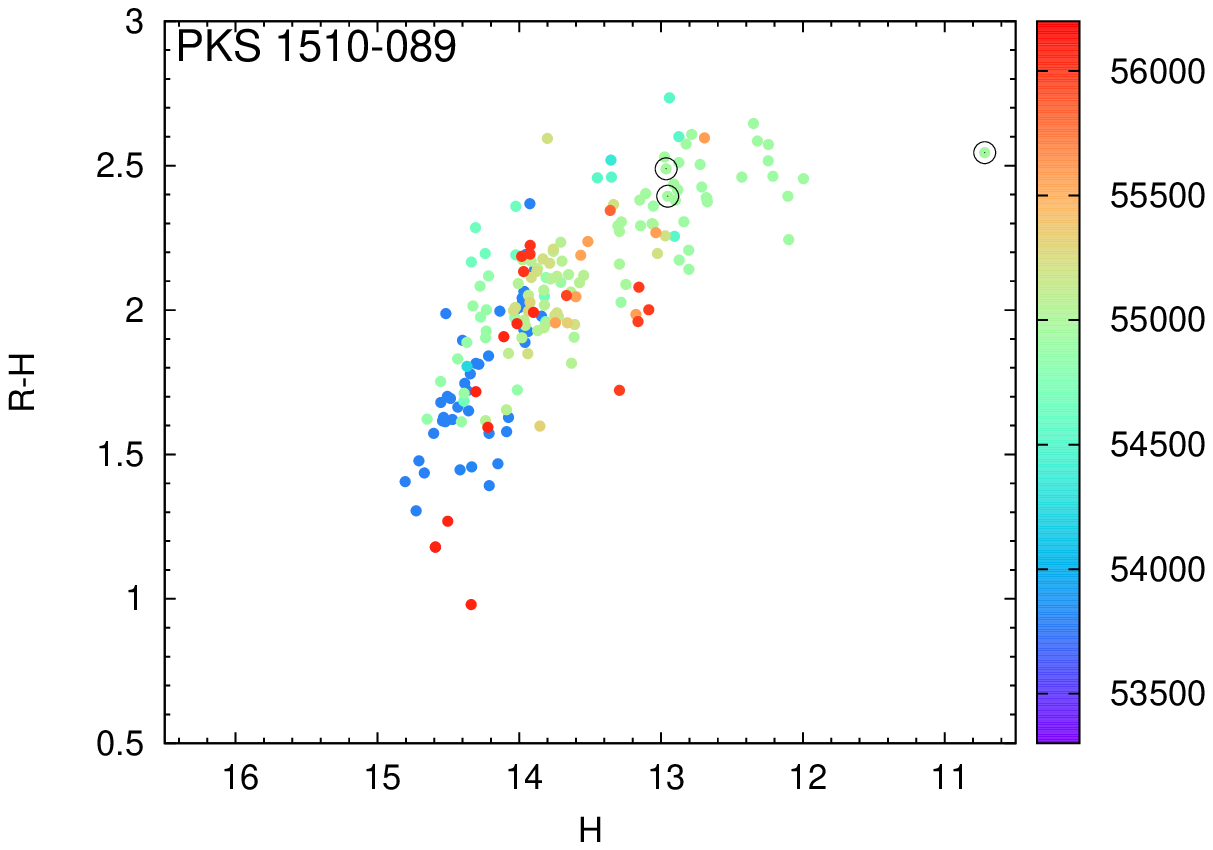}
\caption{Color-intensity diagram for PKS 1510-089. The color of each point is related to the epoch of detection, according to the bar at the right. The circles encompass the points referring  to the flare observed around 54961 MJD. Error bars are omitted for readability.}
\label{1510col}
\end{figure}

\begin{figure*}
 \centering
  \begin{tabular}{cc}
    \includegraphics[width=1\columnwidth]{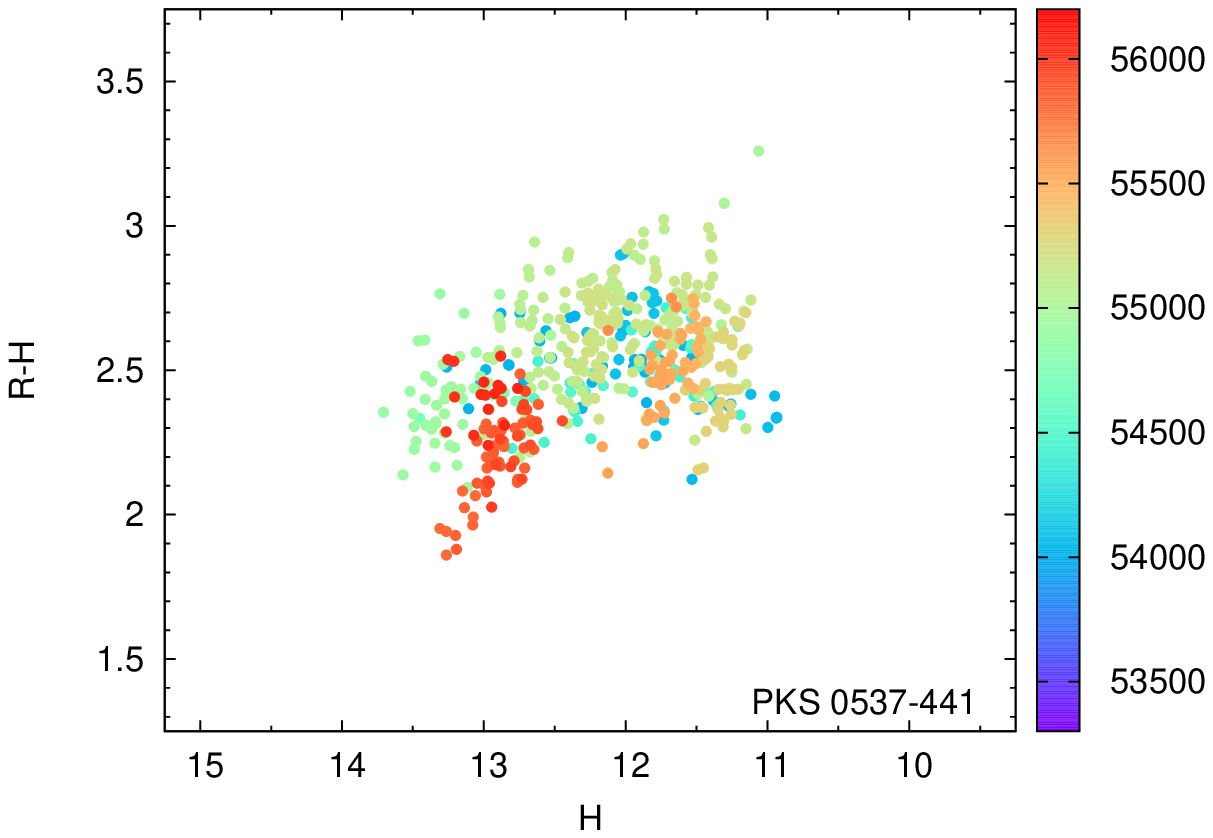}&
    \includegraphics[width=1\columnwidth]{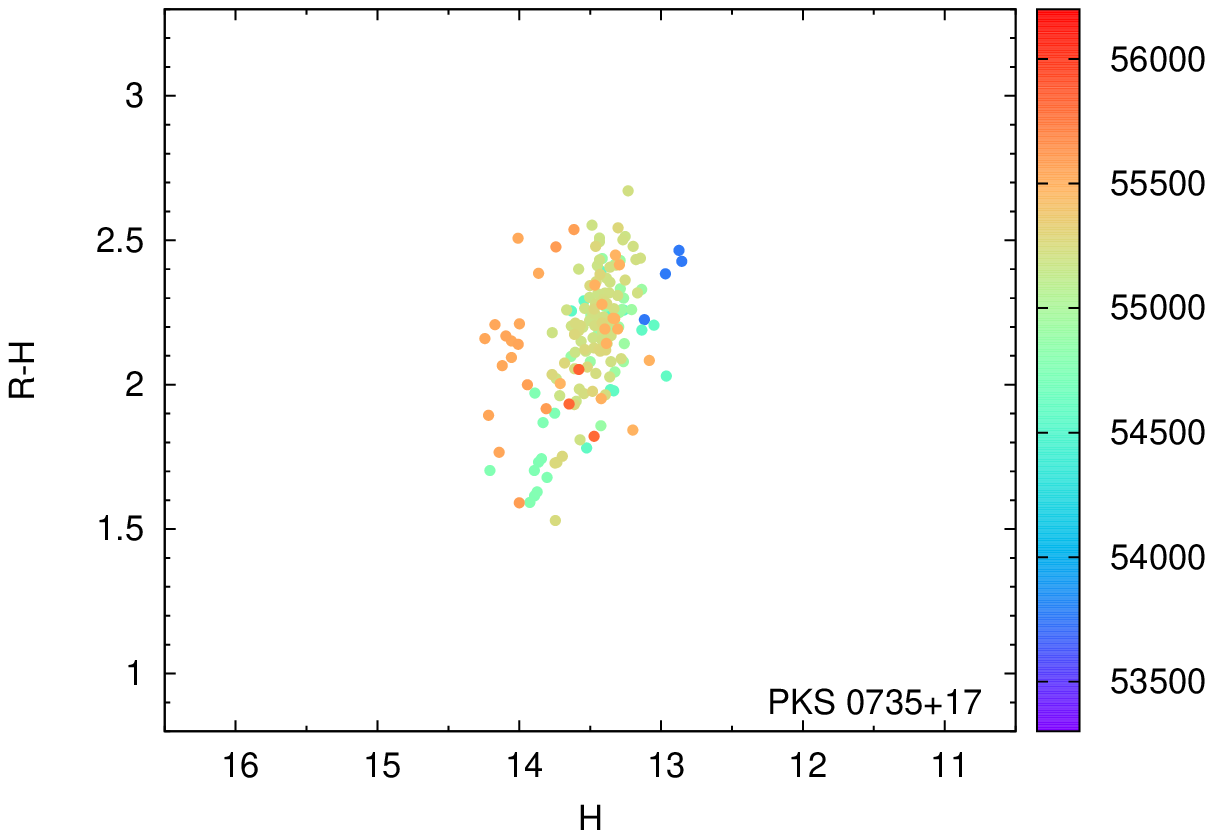}\\
    \includegraphics[width=1\columnwidth] {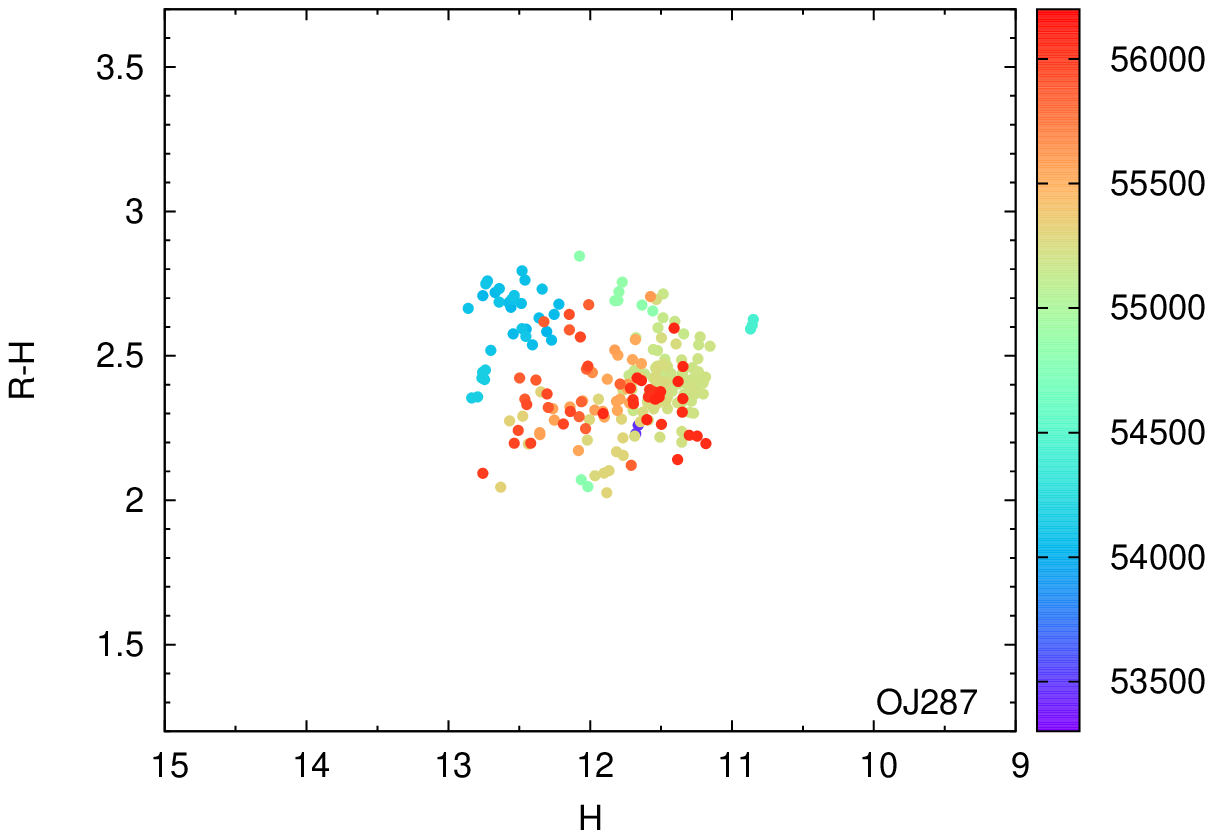}&
    \includegraphics[width=1\columnwidth]{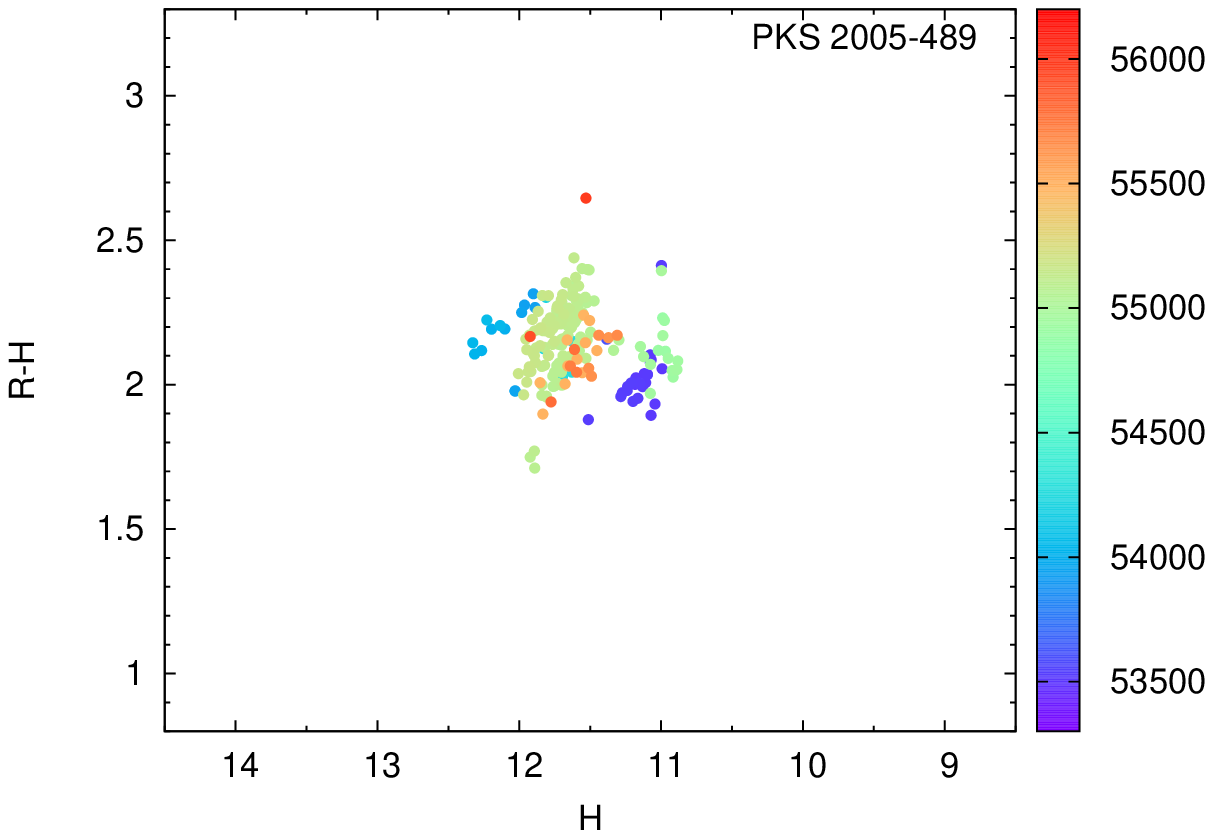}\\
    \includegraphics[width=1\columnwidth]{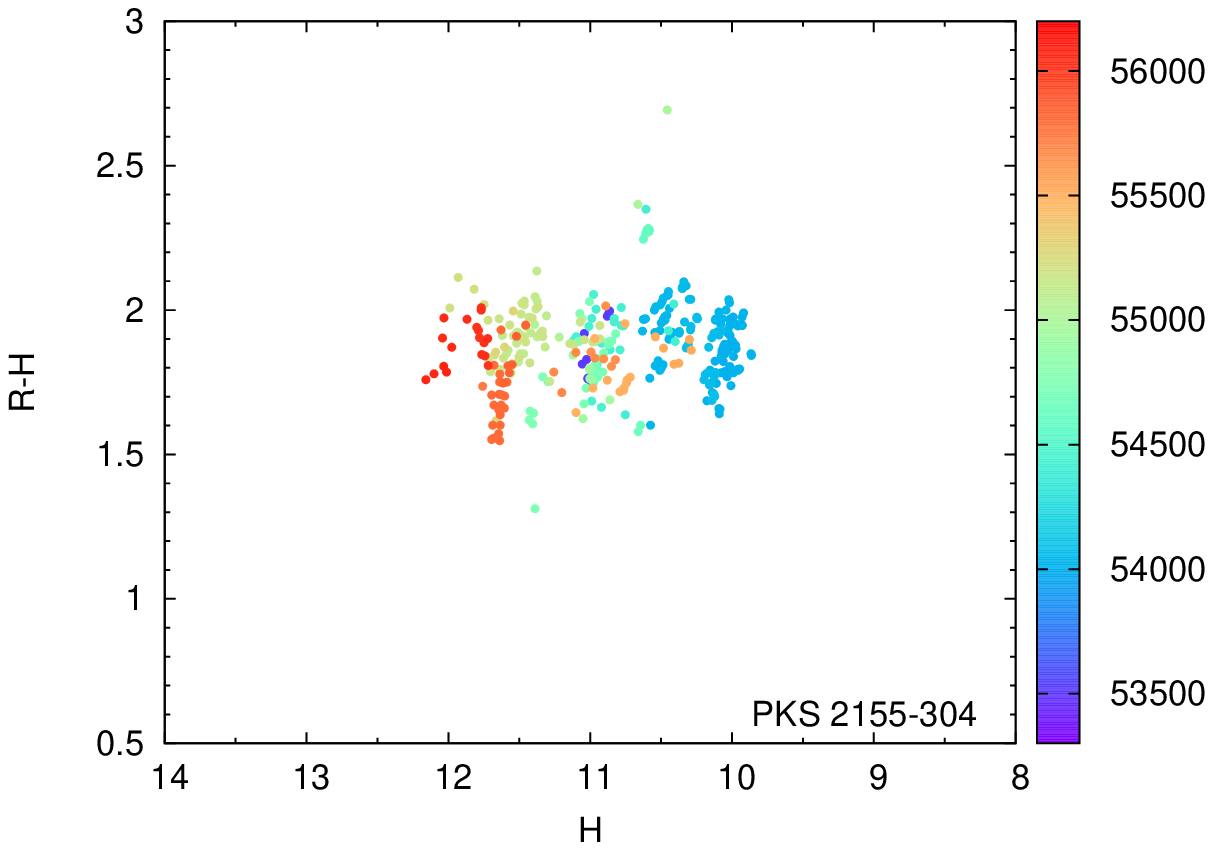}&
    \includegraphics[width=1\columnwidth]{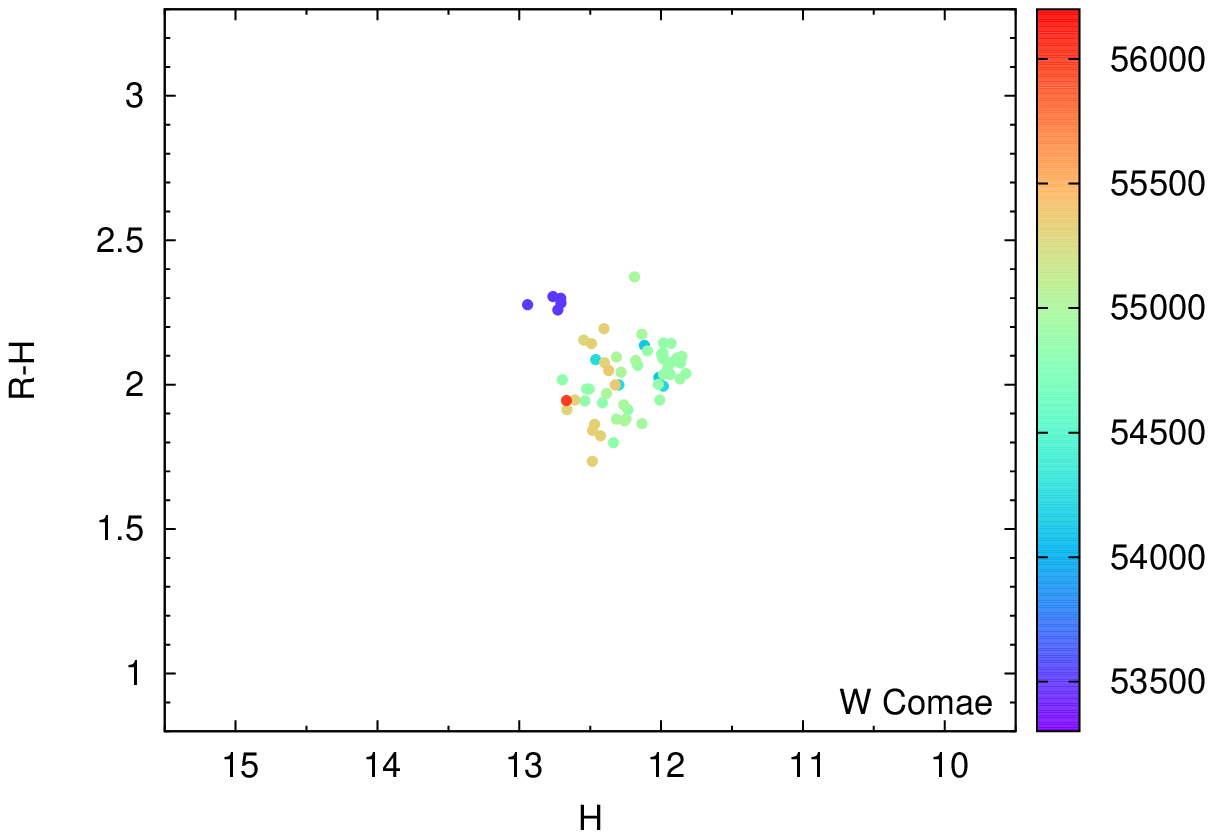}\\
  \end{tabular}
  \caption{Color-intensity diagrams for the BL Lacs. The color of each point is related to the epoch of detections of corresponding data, according to the bar at the right. Error bars are omitted for readability.}
    \label{BLcol} 
    \end{figure*}

For each source the general behavior of the light curves is similar in the various filters. Nevertheless, color-intensity plots are of some interest. We selected  epochs when the time lapse between R and H observations was less than six minutes.  The R-H  color vs the H magnitude is reported in Figs. \ref{1510col} and \ref{BLcol}. It  is apparent that the plots are rather different.  PKS 1510-089 exhibits a banana-like  shape, with a general trend indicating bluer color for decreasing flux.  The same tendency appears in PKS 0537-441, and in PKS 0735+17. On the other hand, OJ 287 and PKS 2005-489 trace a circular spot, or rather an atoll shape.  In the case of PKS 2155-304 the spot is distorted in  the intensity direction.  The shape of the color-intensity  plots may be very different if one only considers a fraction of the overall collection of data (see the bars at the right of the figures).

The case of the flare of PKS 1510-089 around 54961 MJD marked  in Fig.\ref{1510col}  is illustrative of the complexity of the color-intensity dependence. During both the rising and falling phases of the flare, R-H remained constant and  \textit{J-K} was decreasing by $\sim$0.5, indicating a bluer color for increasing intensity. This result may be relevant in order to reconcile the achromaticity of the flare reported by \cite{Bonning2012} on the basis of the \textit{B-J} color \textit{vs J}, and the suggestion  by \cite{Sasada2011} of bluer values for higher intensity indicated by the  \textit{V-J vs V}  plot.

The availability of six filters allows the construction of the spectral flux distributions and the SEDs that are reported in Fig.\ref{1510SED}. In all sources the highest states are well fitted by a single power law, and deviations may appear in the lowest states.

 \setcounter{figure}{6} 
      \begin{figure*}
     \centering
         \includegraphics[width=.45\textwidth]{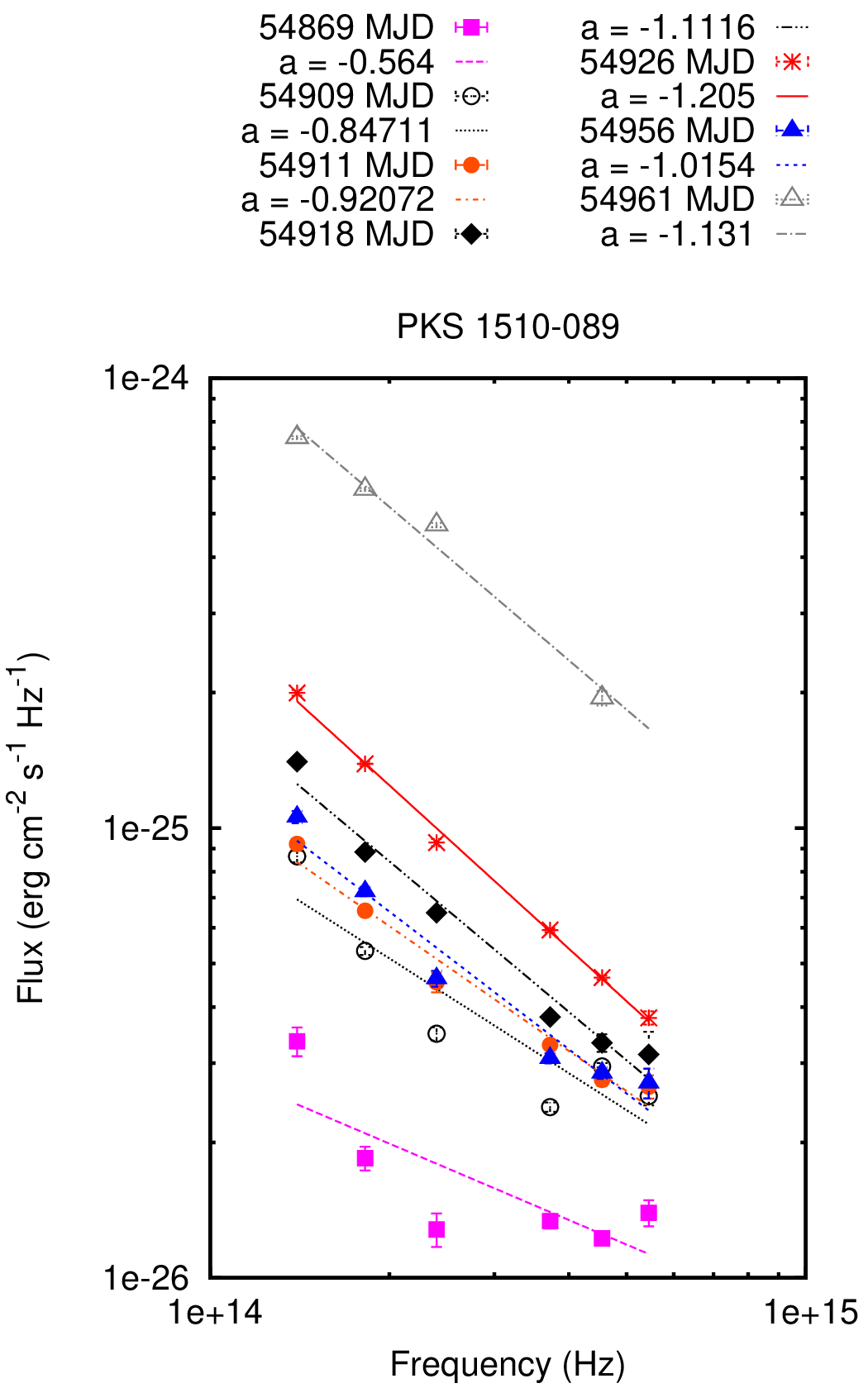}
\includegraphics[width=.45\textwidth]{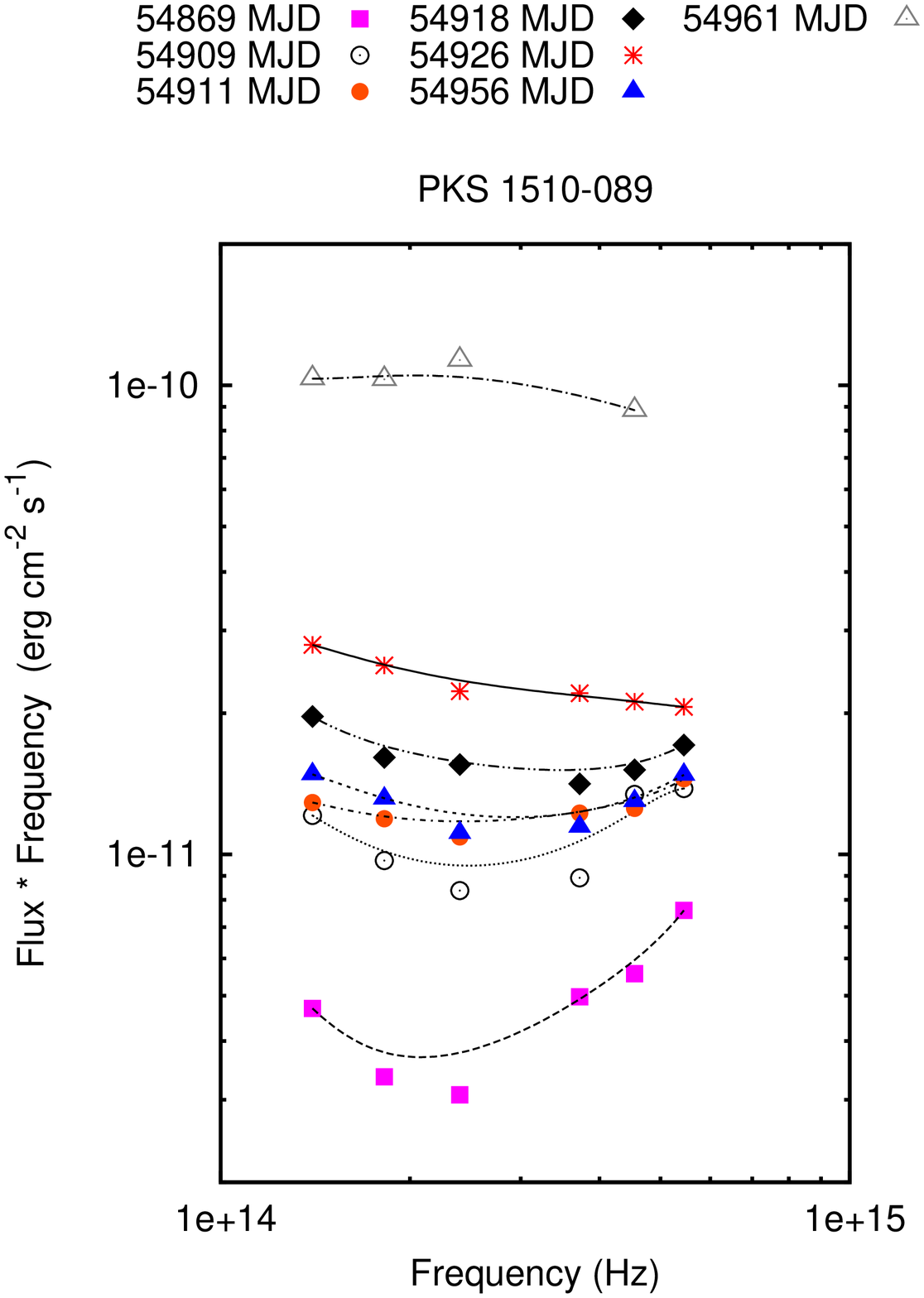}  
 \caption{Spectral flux distribution (left panel) and spectral energy distribution (right panel). The lines in the left panel are power-law fits and corresponding slopes \textit{a} are reported. In the right panel the lines are a help for the eye.}
        \label{1510SED}   
        \end{figure*}

 \setcounter{figure}{6}
   \begin{figure*}
 \centering
 \begin{minipage}[top]{.4\textwidth}
  \centering
 \vspace{1cm}
 \includegraphics[width=\textwidth]{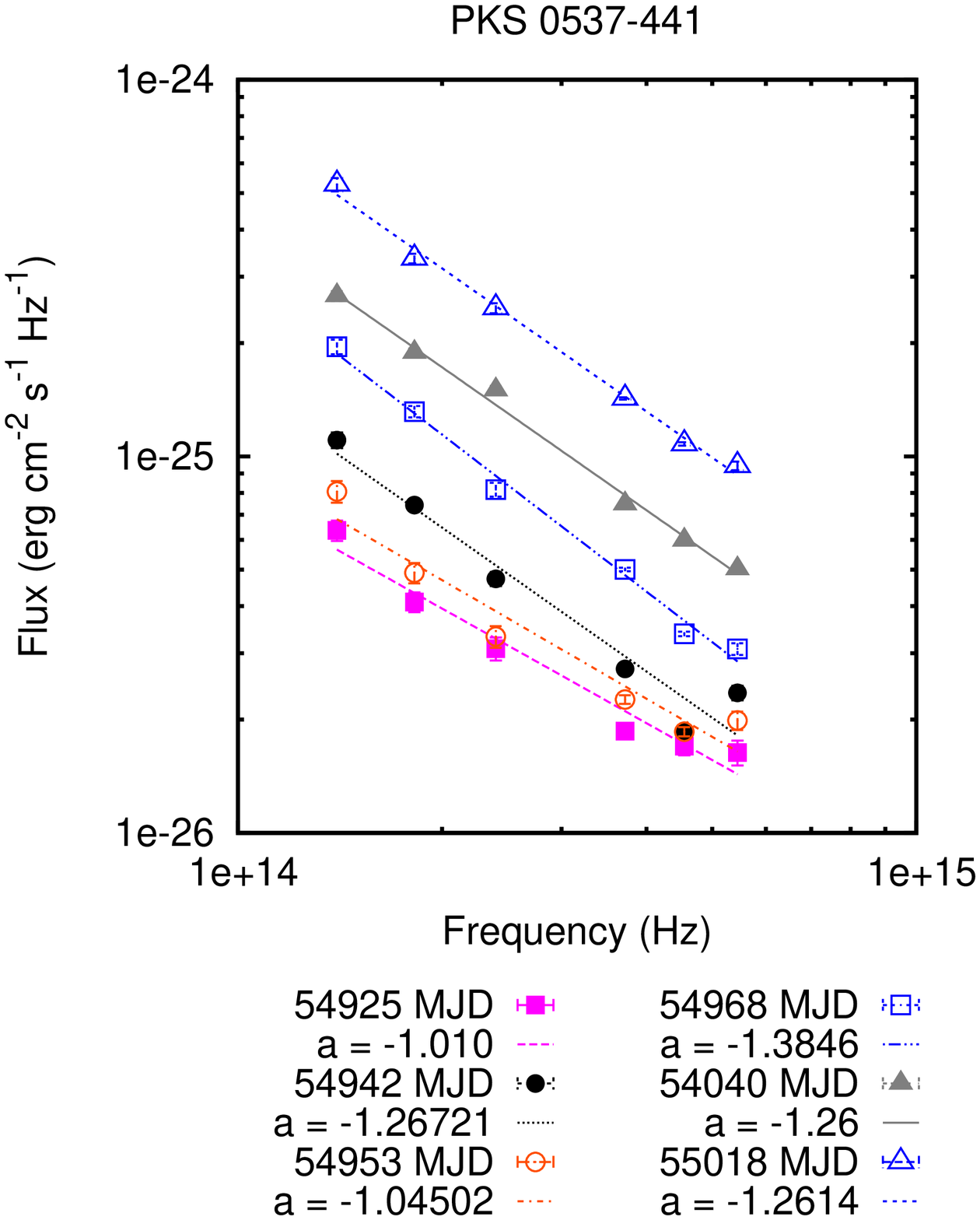}
 \end{minipage}
 \begin{minipage}[top]{.4\textwidth}
  \centering
 \vspace{0cm} 
\includegraphics[width=\textwidth]{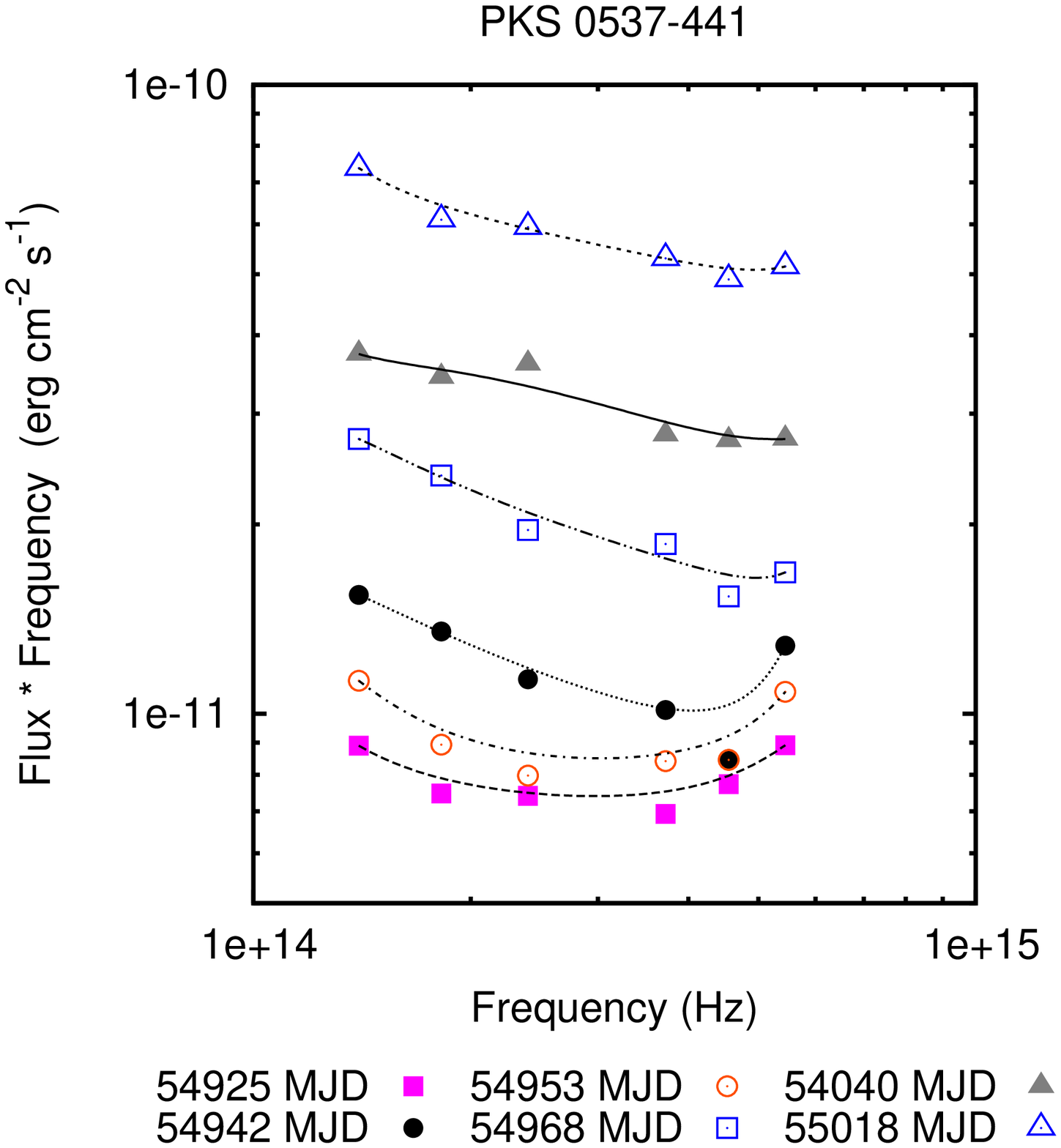}
 \end{minipage}
 \begin{minipage}[top]{.4\textwidth}
  \centering
 \vspace{1cm}
 \includegraphics[width=\textwidth]{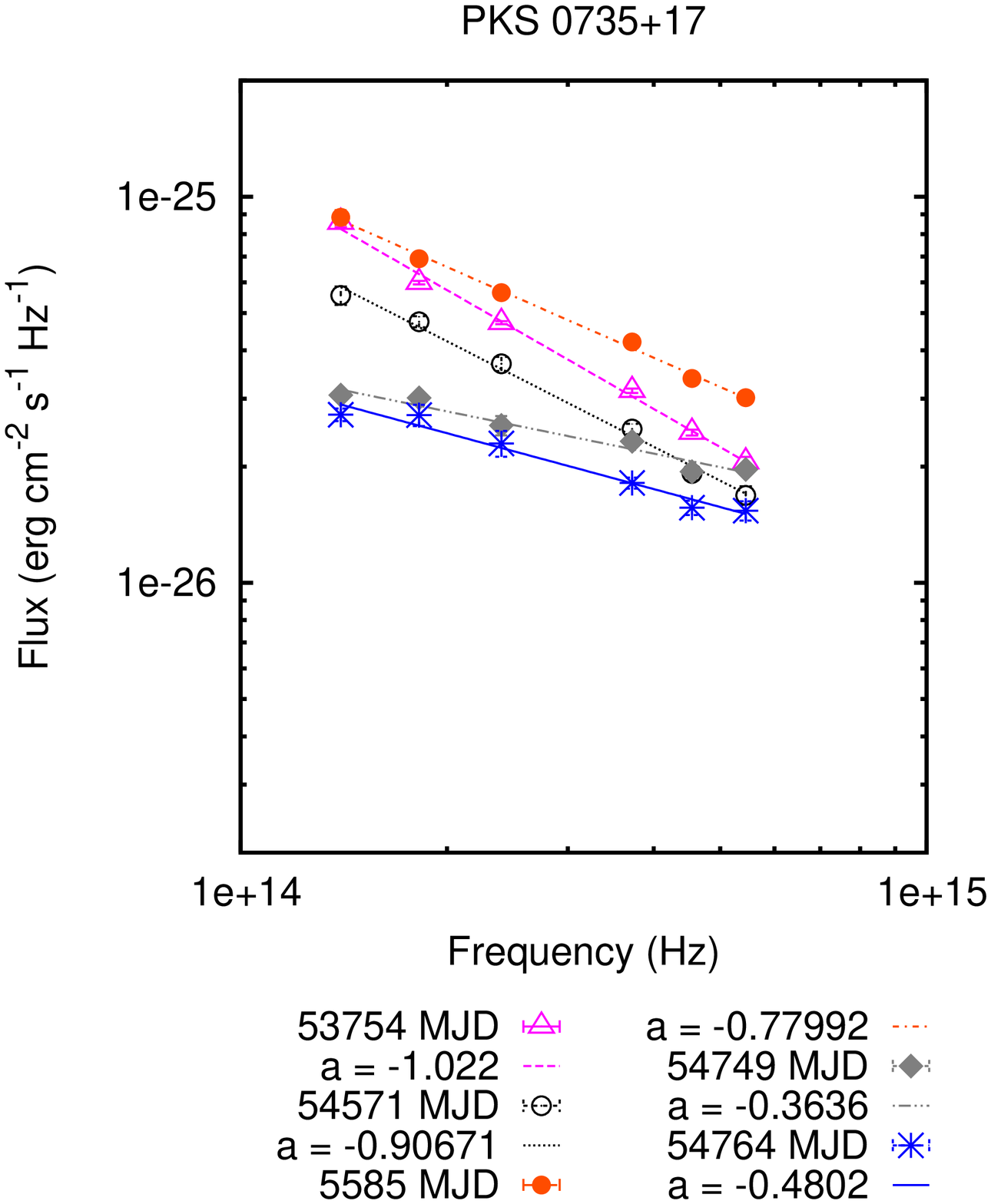}
  \end{minipage}
 \begin{minipage}[top]{.4\textwidth}
  \centering
 \vspace{0cm} 
\includegraphics[width=\textwidth]{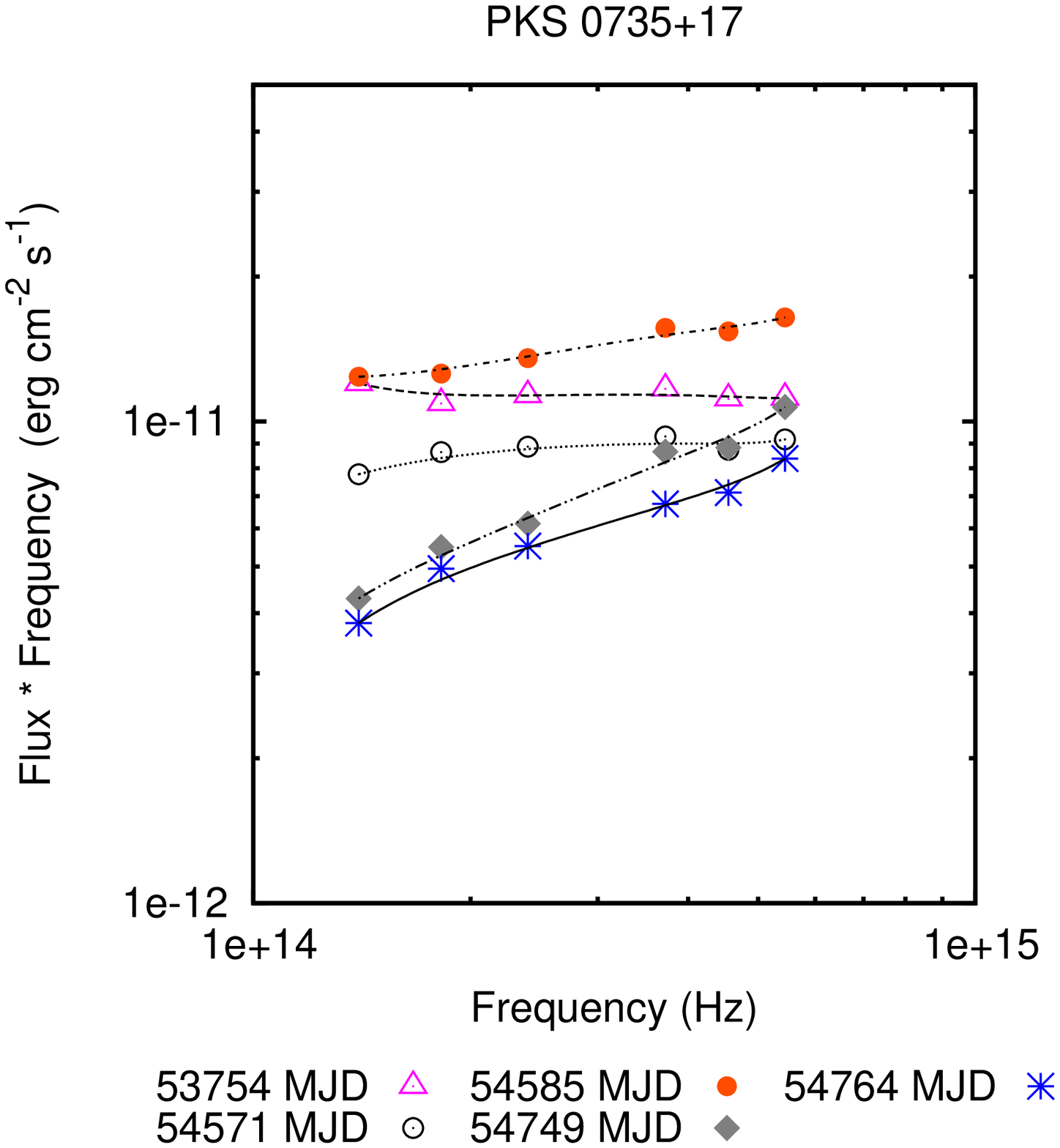}
 \end{minipage}
    \caption{--- Continued}
  \label{0537+0735SED}
          \end{figure*}

 \setcounter{figure}{6}        
 \begin{figure*}
 \centering
 \begin{minipage}[top]{.4\textwidth}
  \centering
 \vspace{1cm}
 \includegraphics[width=\textwidth]{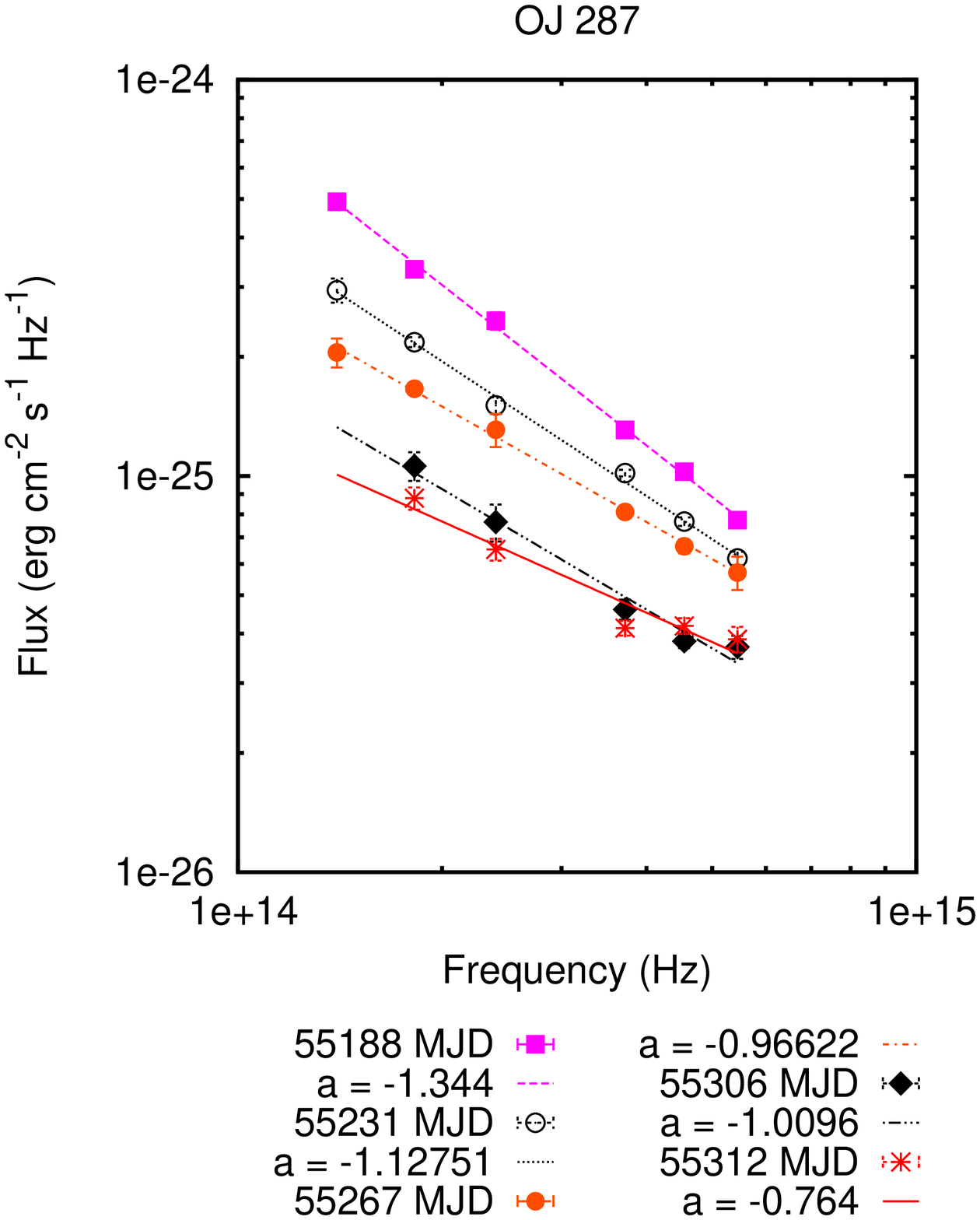}
 \end{minipage}
 \begin{minipage}[top]{.4\textwidth}
 \centering
 \vspace{0cm} 
\includegraphics[width=\textwidth]{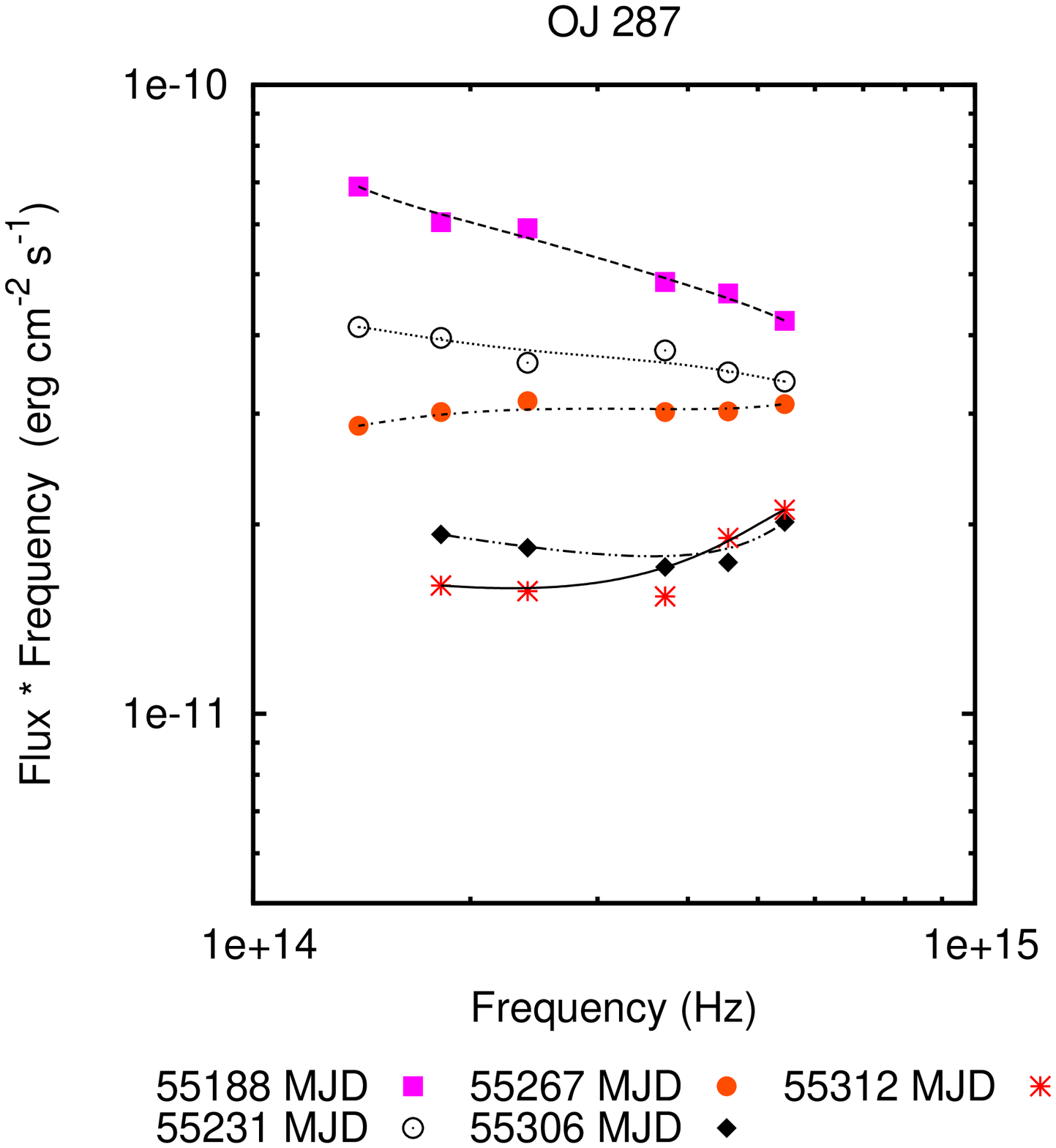}
 \end{minipage}
 \begin{minipage}[top]{.4\textwidth}
   \centering
 \vspace{1cm}
 \includegraphics[width=\textwidth]{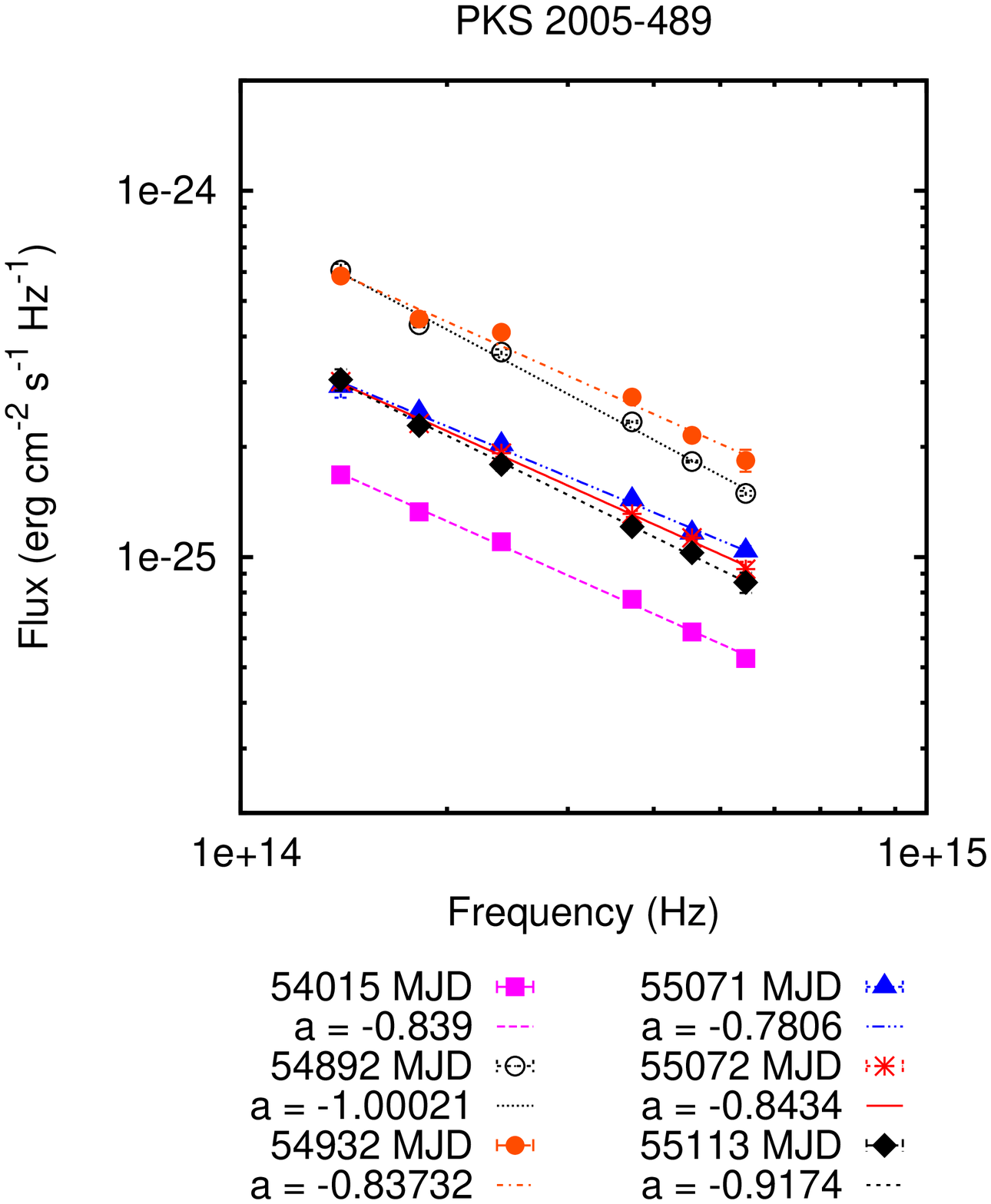}
 \end{minipage}
 \begin{minipage}[top]{.4\textwidth}
  \centering
 \vspace{0cm} 
\includegraphics[width=\textwidth]{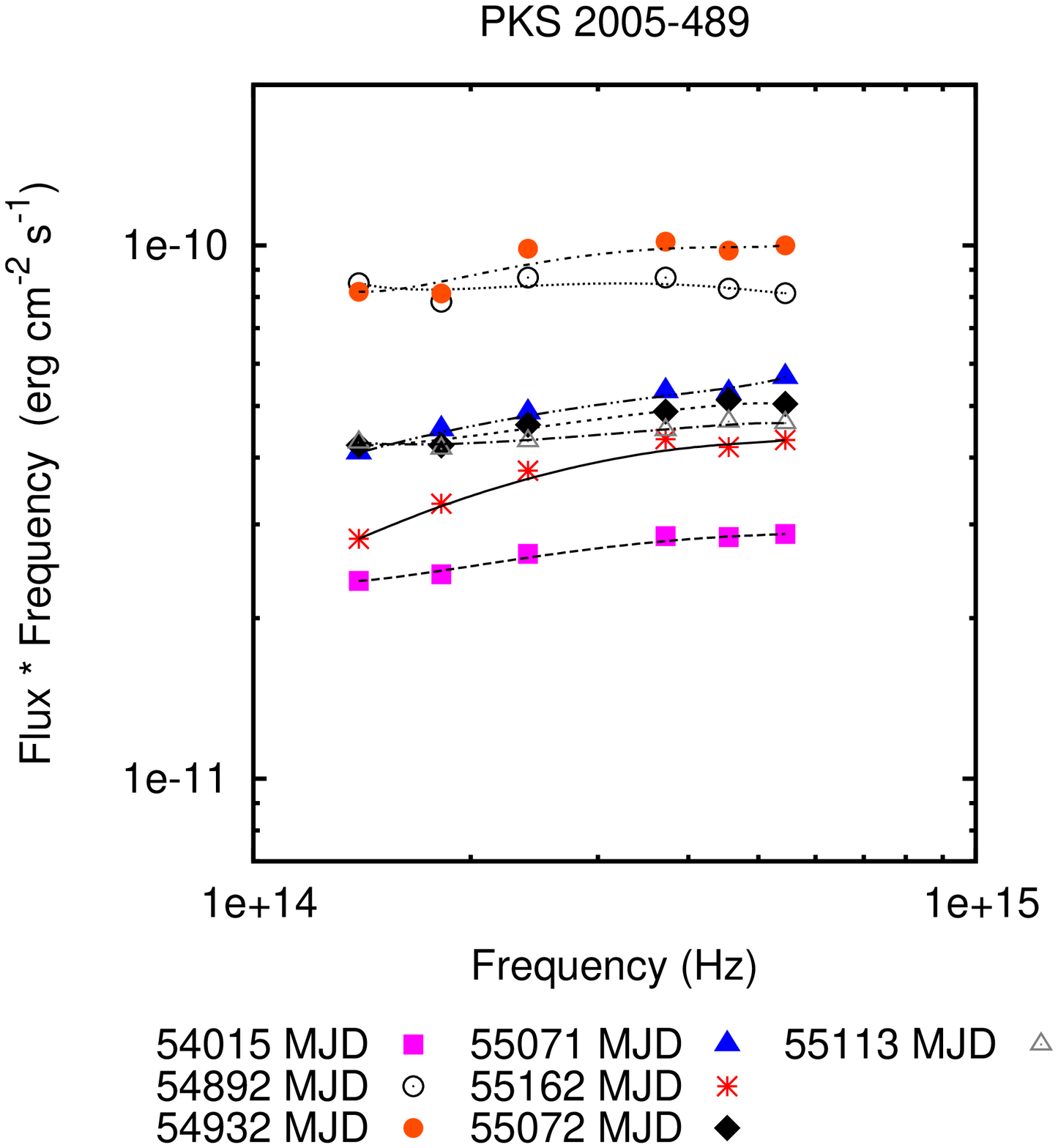}
 \end{minipage} 
  \caption{--- Continued}
  \label{OJ+2005SED}
  \end{figure*}

 \setcounter{figure}{6}  
  \begin{figure*}
 \centering
 \begin{minipage}[top]{.4\textwidth}
  \centering
 \vspace{1cm}
 \includegraphics[width=\textwidth]{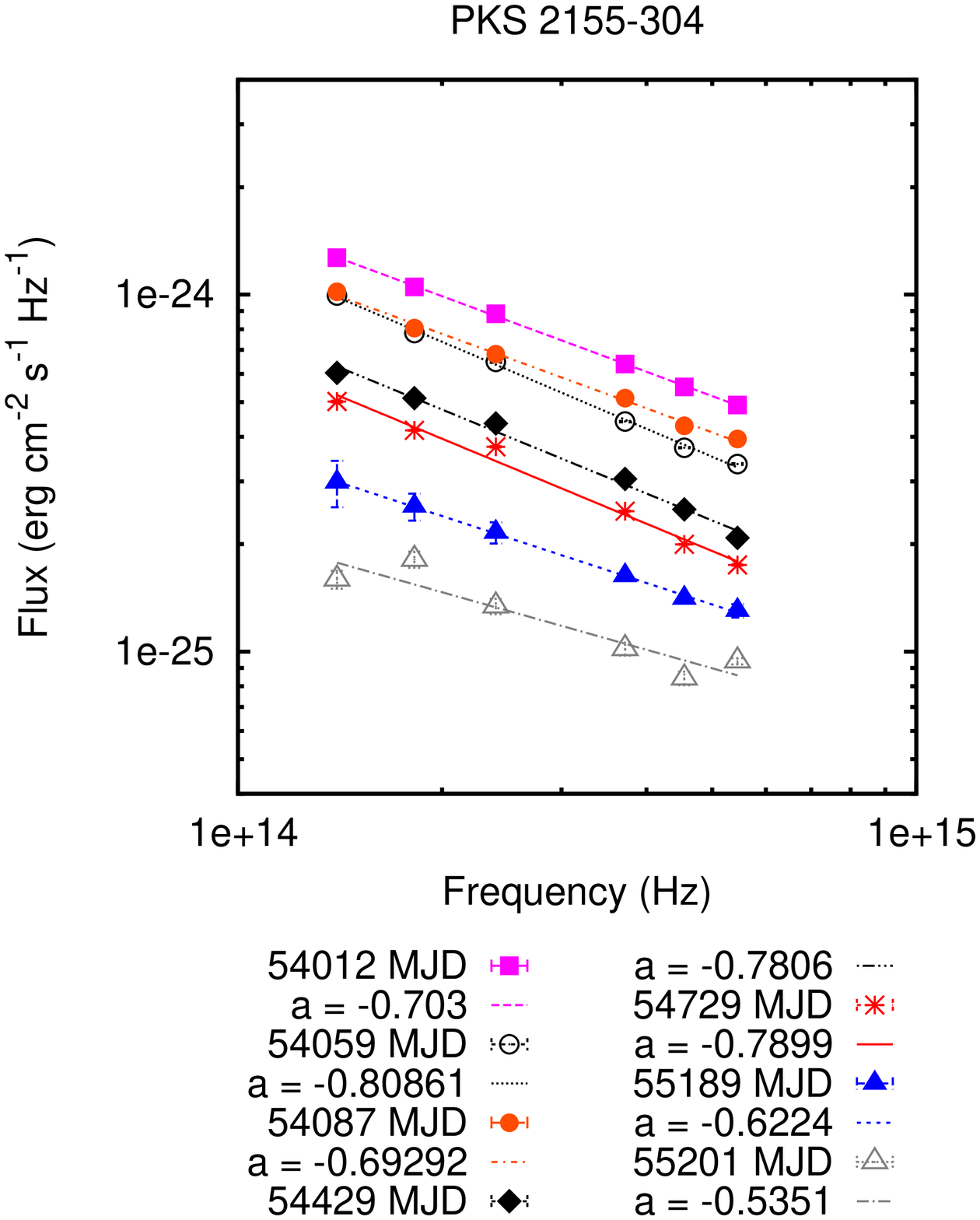}
 \end{minipage}
 \begin{minipage}[top]{.4\textwidth}
  \centering
 \vspace{0cm} 
\includegraphics[width=\textwidth]{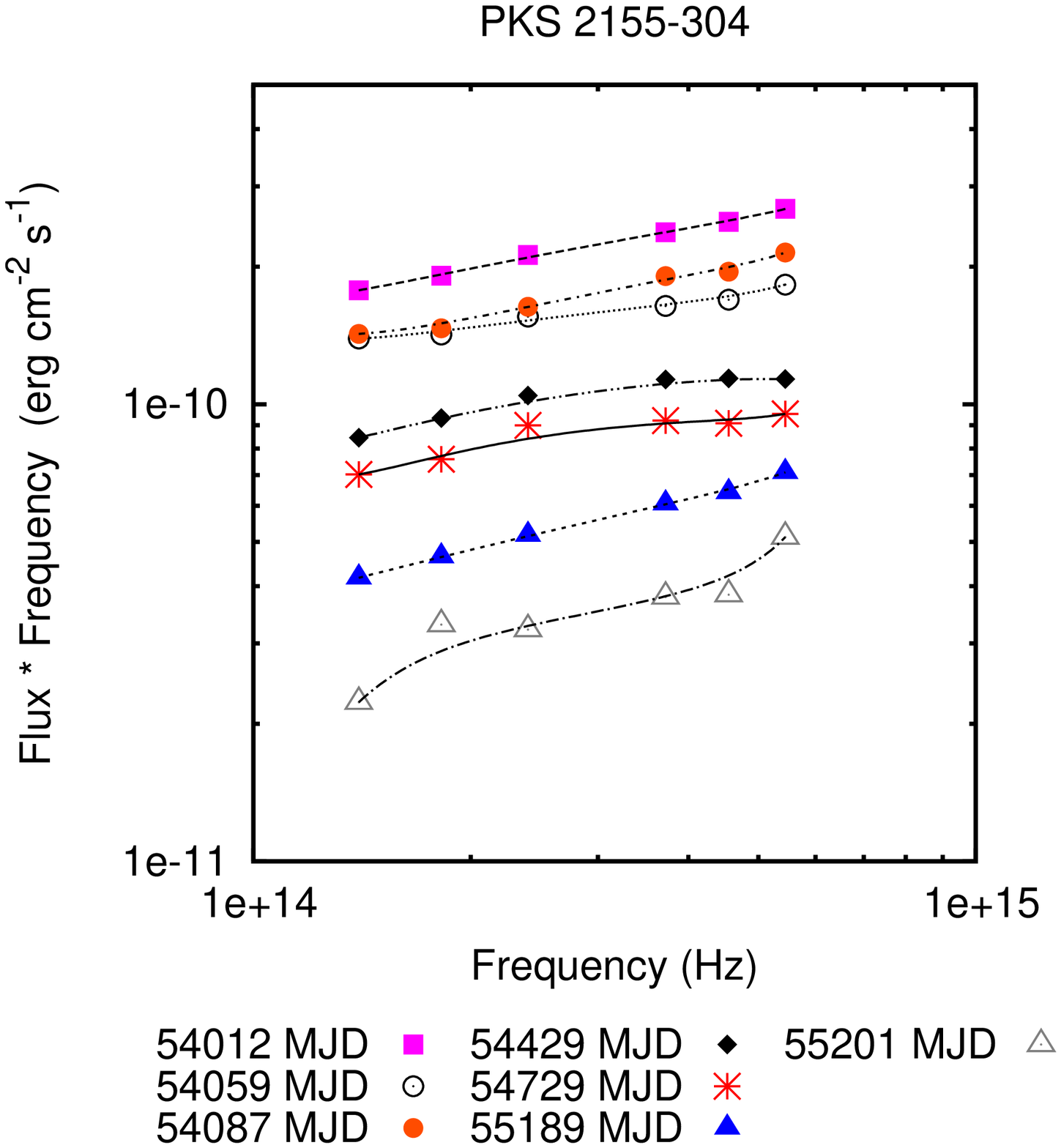} 
\end{minipage}
 \begin{minipage}[top]{.4\textwidth}
  \centering
 \vspace{1cm}
 \includegraphics[width=\textwidth]{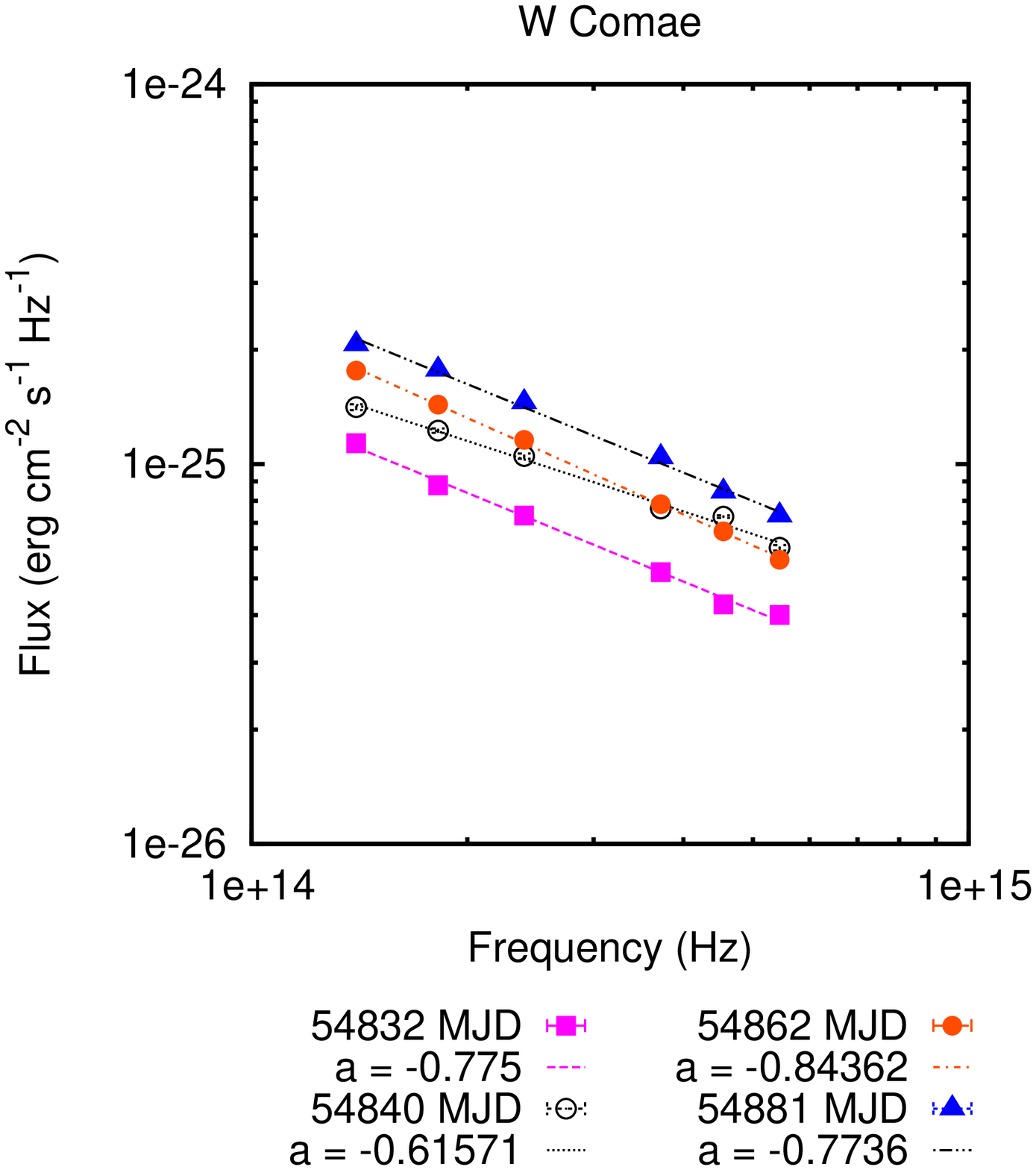}
 \end{minipage}
 \begin{minipage}[top]{.4\textwidth}
  \centering
 \vspace{0.3cm} 
\includegraphics[width=\textwidth]{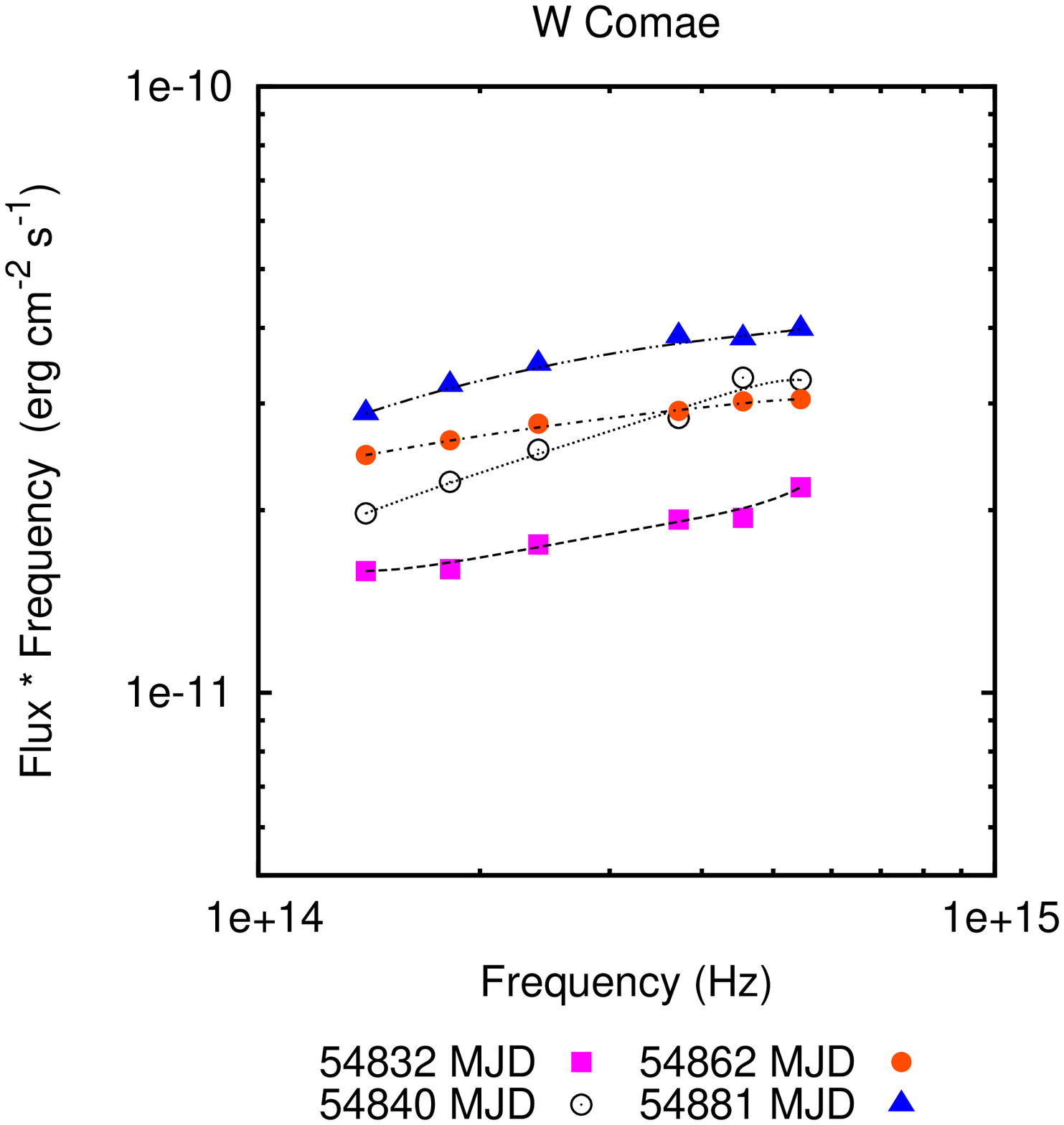}
 \end{minipage} 
 \caption{--- Continued}
  \label{2155+WCSED}
  \end{figure*}


%
\section{Rapid events in the light curves}

REM observations of blazars, and specifically of the seven sources considered here, contain data deriving from proposals with various scientific goals. A large fraction of the observations were performed within multiwavelength campaigns, where the optical and NIR monitoring was ancillary to X-ray  or gamma ray (GeV, TeV) programs. In these cases the rate of monitoring could be typically once a day, or even once a week. The data suitable to search for variability on a time scale of hours are limited to a few epochs, while on much shorter time scales, the constraint is  on the poor count statistics, owing to the diameter of the REM mirror. The threshold on the minimum flux detectable with a given integration time also depends on the mirror reflectance status, which has worsened after 2009, until a general maintenance intervention in 2012. 
When we were searching for rapid episodes, in addition to the visual control of each frame (Section 2),  we chose to consider only those states over a  minimum flux $f_{min}$
 $\sim$  3 mJy  in the optical bands, and  $\simeq$ 1mJy  in the NIR bands.  For fluxes <  $f_{min}$, no search of rapid variability events was performed. To identify potentially relevant variability events, we followed a procedure 
analogous to the one described by \cite{Montagni2006}. We considered a series of N consecutive observations, for each of which the flux is measured.  The flux-time dependence is best fitted with a linear relation, which yields a value of $df/dt$. A time scale is defined as

\begin{equation}
\tau= \frac{<f>}{df/dt} {\frac{1}{(1+z)}}
\label{TS}
\end{equation}\\
where $<f>$ is the average value of the flux and $z$  the redshift. The search process for rapid events is fully automatized. An "event" is selected when $\tau$ is less than a given $\tau_0$, and the uncertainty on $\tau$
is less than $\tau/2$. The event is discarded if 
the check star is significantly variable, and whenever $\tau$ is equal to or larger than $\xi^{-1}\tau_{ck}$,
where $\tau_{ck}$ is the variability time scale of the check star, and $\xi$ is a fixed value. All the images  of an event are then visually examined, excluding the cases where the source or the reference stars are near the image borders or cases where spurious tracks
or spots are apparent. 

\begin{figure*}
   \centering
       \includegraphics[angle=-90,width=1.2\columnwidth] {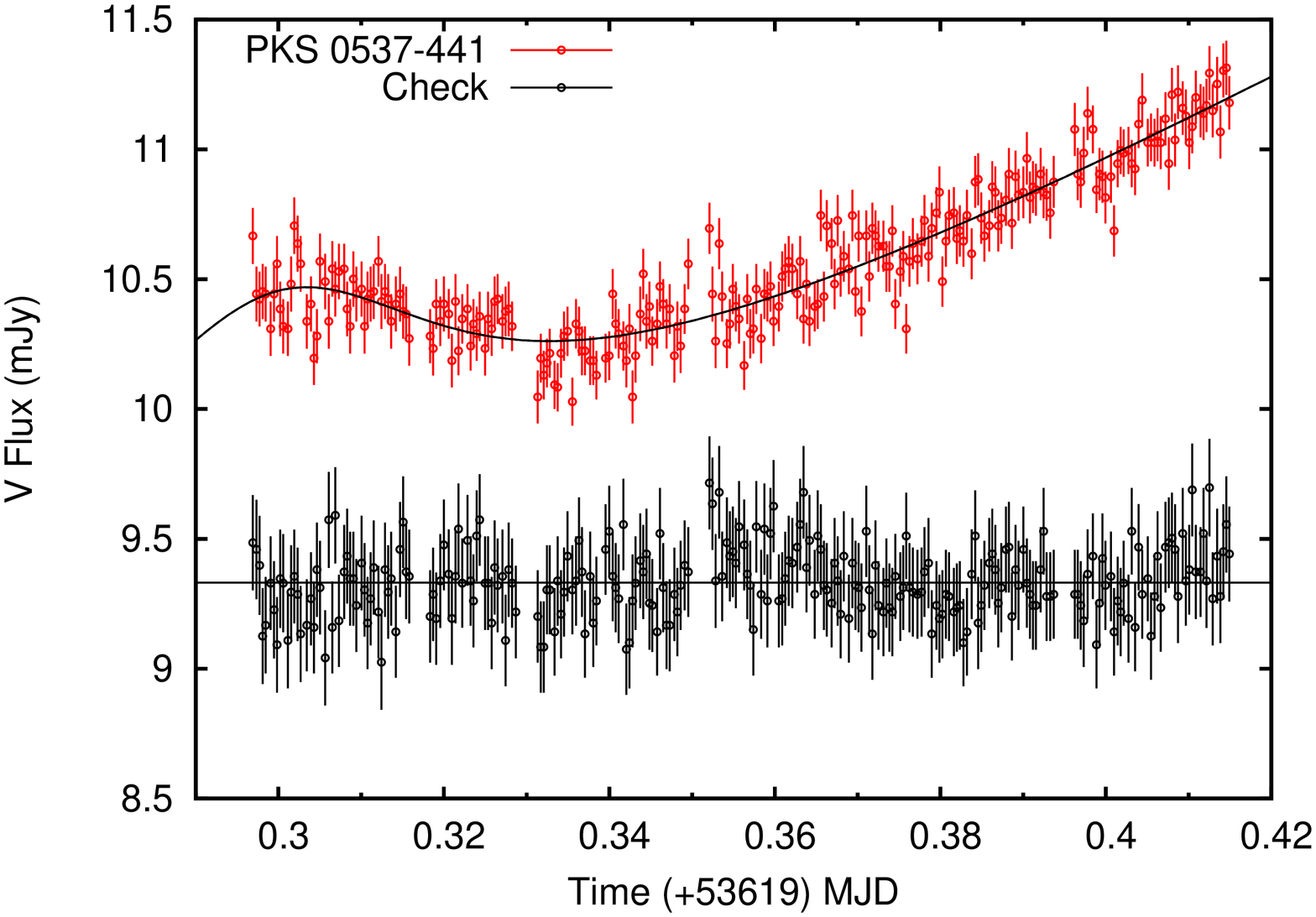}
        \includegraphics[angle=-90,width=1.2\columnwidth]{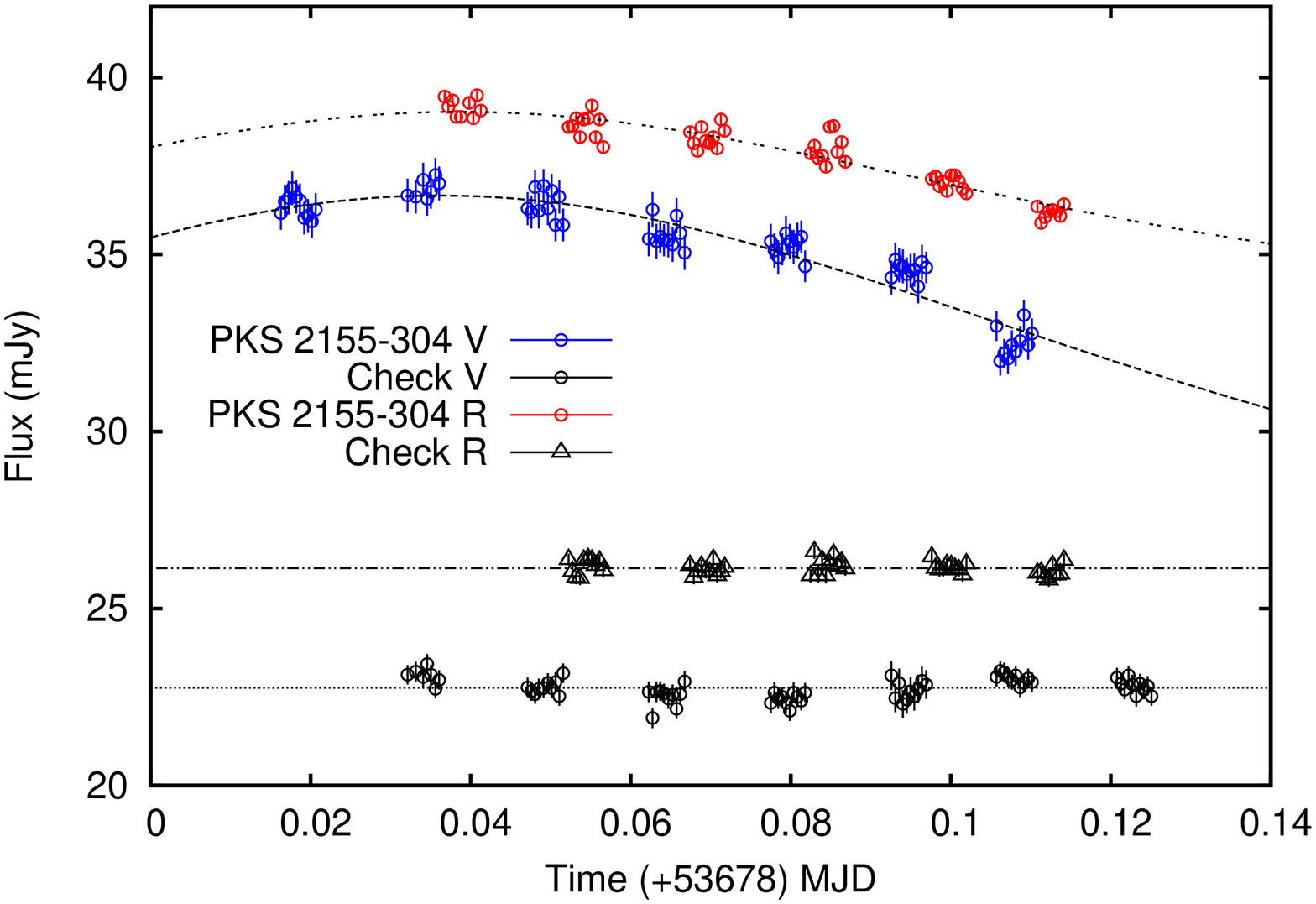}
      \caption{Episodes of intra-night variability in PKS 0537-441and in PKS 2155-304. The check data has been multiplied  x 3 for  PKS 2155-304 and x 2 for PKS 0537-441.The lines are the best fits using eq.\ref{TSExpFit}. }
         \label{RapEv}
   \end{figure*}

First we looked for relatively long  and well documented events, considering $N\geqslant30$ in the same night, with a value of $\tau < 12$ hours and $\xi=10$. We have retrieved two events of this type. The one referring to PKS 0537-441 is reported in Fig.\ref{RapEv}, see also Table \ref{RapEvTab}.  The linear time scale is $\sim$12 hrs.

Following the procedure suggested by \cite{Abdo2010} for Fermi gamma rays sources and adopted by \cite{Danforth2013} for UV photometry of the BL Lac object S5 0716+714, we also fitted the points in the night of the event with the function

\begin{equation}
f(t)=f_0+a \left( e^{(b-t)/\tau_r}+e^{(t-b)/\tau_d}\right)^{-1} 
\label{TSExpFit}
\end{equation}\\
where $f_0$ is a constant underlying flux level, $a$ is a measure of the amplitude of the event,  $b$ roughly describes  the time of the peak of the event, and $\tau$ parameters are the rise and  decay timescales. The  best fit is given in Fig.\ref{RapEv},  and the relevant timescales  are $\tau_r \sim$ 1.5 hr and $\tau_d \sim$ 12 min (see Table \ref{RapEvTab}).  Unfortunately  during the night  of the event, only V photometry was collected. The event has some similarity with a variability episode of PKS 0537-441 deriving from the analysis of the REM archives, which is reported  by \cite{Impiombato2011} and which was discussed in some detail by \cite{Zhang2013}. However a careful analysis  of the CCD images has shown the presence of  a dark spot on the camera focal plane making the reliability of the event dubious, which therefore here is ignored. 

The second event  refers to PKS 2155-304 (53678 MJD),  see Fig.\ref{RapEv}  and Table \ref{RapEvTab}. The event  was  discovered with the automatic procedure in the V band, and has
a very similar counterpart in the R band. No rapid event was detected in NIR bands where coverage is poorer. The continuous lines in Fig.\ref{RapEv} are the fits with exponentials following  \cite{Abdo2010}.

\begin{figure*}
   \centering
   \includegraphics[angle=-90,width=0.7\columnwidth]{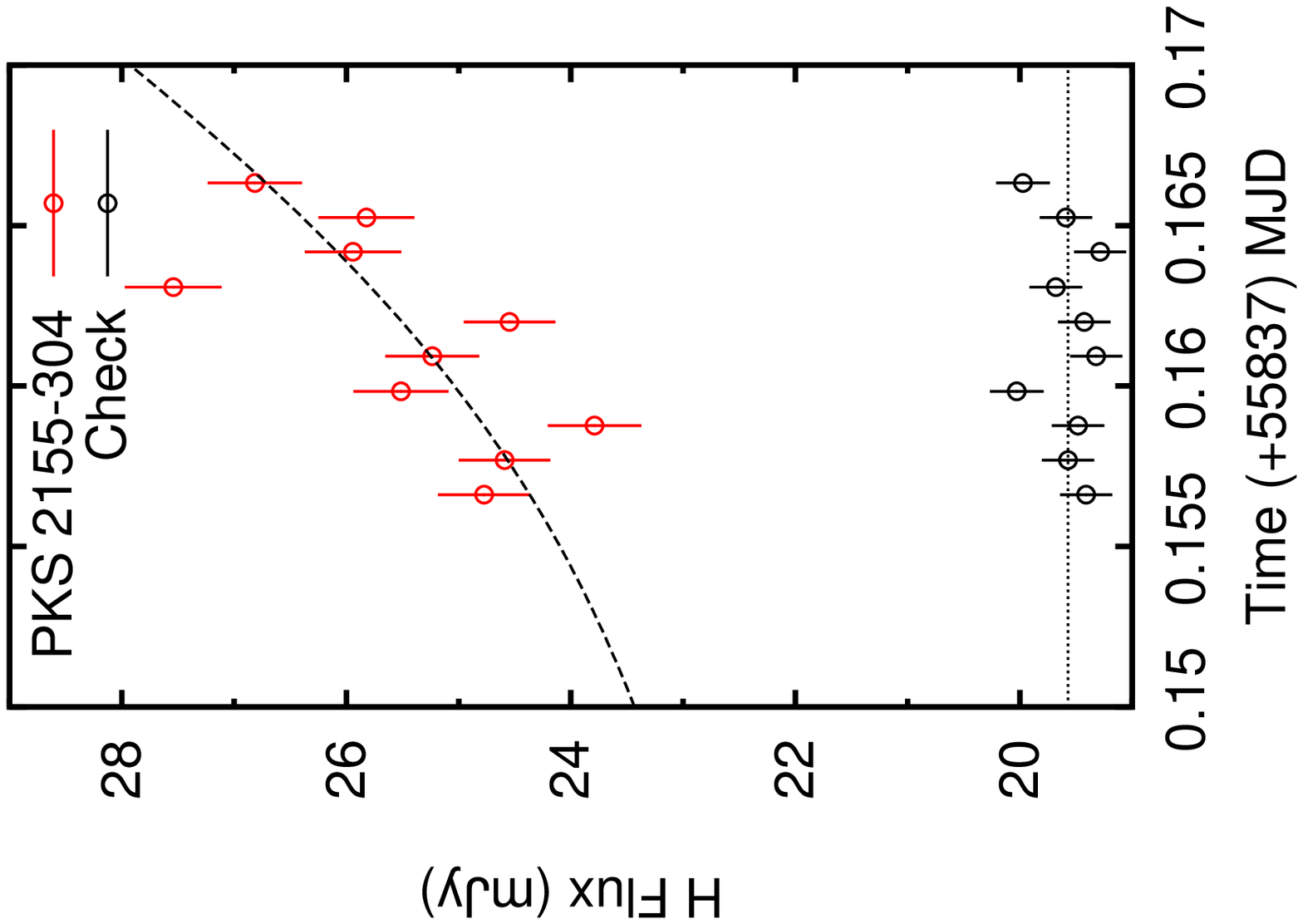}
      \includegraphics[angle=-90,width=0.7\columnwidth]{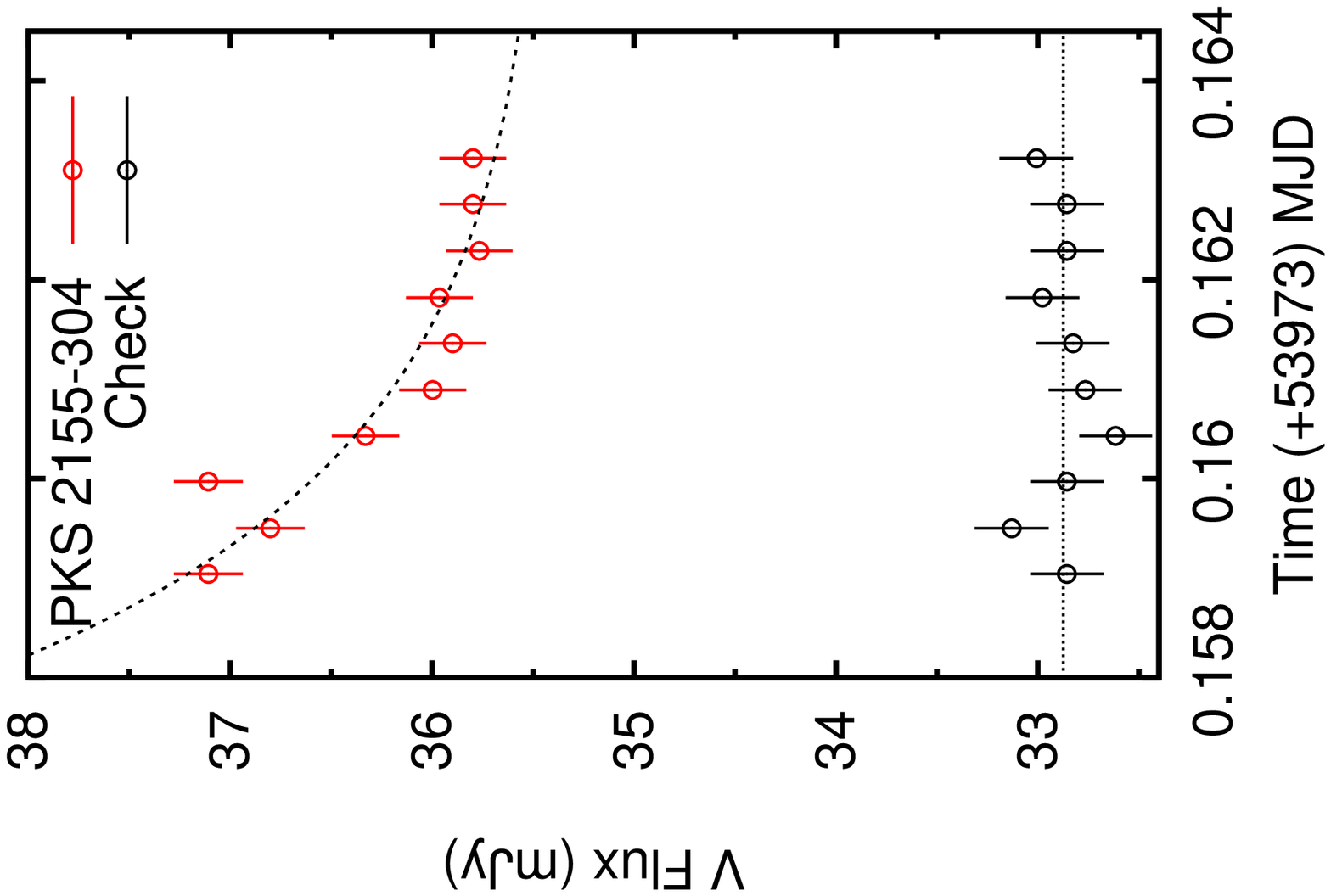}
      \caption{Short variability events. In the right panel the check data has been divided by 2.5. In the left panel the check data has been multiplied  x 1.4. The lines are the best fits using eq.\ref{TSExpFit}.}
         \label{RapRapEv}
   \end{figure*}

We  then looked for more rapid events. We have fixed a time scale $\tau_0< 3h$, $\xi=4$ and a number N=8 of points. The only two events that we retrieved both refer to PKS 2155-304,  see Fig.\ref{RapRapEv} and Table \ref{RapEvTab}. 
The exponential time scales are now of the order of minutes. The event occurred on 53973 MJD yields a time scale of three minutes and the reduced $\chi^2$ indicates an acceptable  fit with an exponential function. The case in 55837 MJD is more complex. It seems that superimposed on the exponential rise there may be a spike that is, however, represented by a single point. We detect for both these episodes the very fast variability rate of 0.43 mag/hr , comparable to the variability rate of 0.45 mag/hr found by  \cite{Danforth2013} for S5 0716+714.

\begin{table*} 
\centering
\scriptsize
\caption{Rapid event time scales.}
\begin{tabular}{lccc|cc|ccc}
 \hline
 \hline
Source			&Filter&MJD	&N	&\multicolumn{2}{|c|}{Linear fitting } 						& \multicolumn{3}{c}{Exponential fitting}							\\
				&	&		&	&$\tau$				&$\tau_{ck}$ 					&$\tau_r$				& $\tau_d$			&reduced $\chi^2$\\
\hline\hline
&&&&&&&&\\
PKS 0537-441		&V	& 53619	&198&11h 53m $\pm$ 16m	&34d 17h 22m $\pm$ 1h 56m 		&1h 27 m $\pm$ 6h  48m	&12m 24s $\pm$ 8m 13s	&1.13	 \\ 
				&	&    		&	&		 			&							&					& 					&	 \\ 

 PKS 2155-304		&H	&55837	& 10 	&2h 15m $\pm$ 22m 	&1d 7h 3m  $\pm$ 22h 16m 		&				 	&22m  $\pm$ 1h 30m	&4.60\\
				&R	&53678	& 25 &11h 54m $\pm$ 25m 	&6d 10h 34m $\pm$ 1d 23h 48m 	&					&1h 21m $\pm$ 39m 	&5.78\\
				&V	& 53678	& 31	&12h 00m $\pm$ 21m 	&7d 16h 28m $\pm$ 3d 17h 44m 	&					&1h 39m $\pm$ 39m  	&1.16	 \\ 
				&V	& 53973	& 10	&2h 15m $\pm$ 14m 	& 3d 17h 33m 25s  $\pm$ 1d 11h 40m&					&3m 14s $\pm$ 2m 1s	&1.79	 \\ 
&&&&&&&&\\
\hline
\end{tabular}
\tablefoot{The exponential fits refer to eq.\ref{TSExpFit}. For the  episodes related to PKS 2155-304, in  eq.\ref{TSExpFit} $\tau_r$ is assumed equal to  $\tau_d$}
\label{RapEvTab}
\end{table*}

\section{Discussion}

For the sources where the SED of the lower states are not well fitted by a power law (PKS 1510-089, PKS 0537-441, OJ 287, see Fig.\ref{1510SED}), the indication is that the NIR bands yield a power-law index, which is unchanged with respect to the high states obtained by combining NIR and optical observations. In the lowest states a new component becomes apparent that shows up in the optical.  This component could be thermal, but this cannot be assessed, owing to the limited bandwidth covered by our six filters. In any case, the second component appears to be bluer than the power law. If the power law represents the variable part, this is consistent with the behavior of the variability index $\sigma_{rms}^{2}$, which tends to be larger in the NIR with respect to the optical bands (see Table \ref{EVave}).

As shown in the previous section we have selected a number of episodes with time scales of hours or even shorter. Our selection criteria have been rather stringent. There is a comparable number of events, which may be real but do not satisfy one of our criteria. All the events derive from two sources  that are very variable on time scales longer than one day, i.e. PKS 0537-441 and PKS 2155-304, and for which the total coverage is the largest, even with regard to the  intra-night monitoring  programs. For the other very variable source, PKS 1510-089, we collected a number of a priori interesting events, but they corresponded to very low states
of the source, and therefore are not considered here for the criterion given at the beginning of Sect.5.  Even when also taking the unconfirmed rapid events into account, one finds that hour time scales are rare. The ratio of their summed duration to the total exposure time  is $\sim$8/670 for PKS 0537-441 and $\sim$ 20/730  for PKS 2155-304, where 670 hrs and 730 hrs are the overall observing times of the two sources.  The discovery of these rapid events therefore requires very long monitoring campaigns. This inference cannot be applied to the case of  S5 0716+714, where the one and only COS-HST observation found a rapid event.

The two most extreme events are those referring to PKS 2155-304 reported in Fig.\ref{RapRapEv}, where the time scale events for a linear fit are on the order of hours, and those for the exponential fit  are a few minutes. Similar short time scales  have been found at  $\lambda$=1400 $\AA$   by \cite{Danforth2013} during the above-mentioned observation of  S5 0716+714. In PKS 2155-304 sub-hour time scales have appeared in the TeV band during the famous active state on July 29-30, 2006 \citep{Aharonian2007}. The mass of the black hole in PKS 2155-304, is most probably $\sim10^{9}  M_{\odot}$ \citep{Falomo1991, Kotilainen1998}, corresponding to a Schwarzschild radius $r_G \sim 2\cdot10^{14}$ cm.  Assuming an emitting region of a few (say 5) $r_G$ , the corresponding time scales are

\begin{equation}
\tau\sim\frac{R (1+z)}{c\delta}=\frac{4\cdot10^{4}s }{\delta}
\label{RelTS}
\end{equation}\\
where $\delta$ is the Doppler factor. A direct comparison with the reported time scales would require a specific model of the source, indicating in particular if the time scale to be considered is from the linear or exponential fit.  In any case the reported rapid optical episodes indicate that the  requirement of large Doppler factors is inevitable, in agreement with what is concluded for the variability at high energies.

\begin{acknowledgements}
      We are grateful to the referee for the very positive and constructive report.

We also  thank Dino Fugazza for his efficient management of the REM observations and archive.

This work has been supported by ASI grant I/004/11/0 and by PRIN-MIUR 2009 grants.    
\end{acknowledgements}

\end{document}